\documentclass[journal,onecolumn]{IEEEtran}
\IEEEoverridecommandlockouts

\usepackage{euscript}
\DeclareMathAlphabet\EuRoman{U}{eur}{m}{n}
\SetMathAlphabet\EuRoman{bold}{U}{eur}{b}{n}
\newcommand{\eurom}{\EuRoman}

\usepackage{bbm}
\usepackage{bm}
\usepackage{amssymb}

\usepackage[utf8]{inputenc} 
\usepackage[T1]{fontenc}
\usepackage{url}
\usepackage{ifthen}
\usepackage[noadjust]{cite}
\usepackage[cmex10]{amsmath}
                             
\usepackage[english]{babel}

\usepackage{mleftright}

\usepackage{tikz}

\usetikzlibrary{math}
\usetikzlibrary{shapes}
\usetikzlibrary{positioning}
\usetikzlibrary{arrows}
\usetikzlibrary{arrows.meta}

\usepackage{pgfplots}
\pgfplotsset{compat=newest}
\pgfplotsset{plot coordinates/math parser=false}
\newlength\figureheight
\newlength\figurewidth

\usepackage{orcidlink}
\usepackage{xcolor}
\usepackage{hyperref}
\usepackage{soul}
\hypersetup{%
  colorlinks = true,
  linkcolor  = red,
  citecolor  = blue,
  urlcolor   = blue
}
\usepackage{soul}
\setul{1pt}{.4pt}

\usepackage[nameinlink]{cleveref}
\Crefname{figure}{}{}

\newcommand{\C}{\eurom{C}}

\newcommand{\G}{G}
\newcommand{\Ell}{\ell}

\newcommand{\I}{\eurom{I}}
\newcommand{\mi}[2]{{\I}\mleft(#1 \, ; #2 \mright)}

\newcommand{\h}{\eurom{h}}
\newcommand{\ent}[1]{{\h}\mleft(#1\mright)}

\newcommand{\Exp}{\eurom{E}}
\newcommand{\expect}[1]{{\Exp}\mleft[#1\mright]}

\newcommand{\varn}{\sigma_z^2}

\newcommand{\Gammaf}[1]{\Gamma \mleft( #1 \mright)}

\newcommand{\f}{f}
\newcommand{\func}{{\f(t)}}
\newcommand{\convf}{\f^*(\theta)}

\newcommand{\Yv}{{\bf Y}}
\newcommand{\Hv}{{\textsf{H}}}
\newcommand{\Xv}{{\bf X}}
\newcommand{\xv}{{\bf x}}
\newcommand{\yv}{{\bf y}}
\newcommand{\Zv}{{\bf Z}}
\newcommand{\Qv}{{\bf Q}}
\newcommand{\Iv}{\textsf{I}}
\newcommand{\Uv}{\textsf{U}}
\newcommand{\QMv}{\textsf{Q}}
\newcommand{\Dv}{\mathsf{\Lambda}}
\newcommand{\Vv}{\textsf{V}}
\newcommand{\Rv}{\textsf{R}}
\newcommand{\Sv}{\textsf{S}}
\newcommand{\rv}{\textbf{r}}
\newcommand{\p}{\textbf{p}}
\newcommand{\sv}{\textsf{s}}
\newcommand{\tYv}{{\bf \widetilde{Y}}}
\newcommand{\tXv}{{\bf \widetilde{X}}}
\newcommand{\tZv}{{\bf \widetilde{Z}}}

\newcommand{\sX}{{\cal X}}
\newcommand{\sF}{{\cal F}}
\newcommand{\sS}{{\cal S}}
\newcommand{\sD}{{\cal D}}

\newcommand{\sSbox}{\sS_A}
\newcommand{\sSball}{\sS_B}

\newcommand{\sR}{{\cal R}}
\newcommand{\sP}{{\cal P}}
\newcommand{\sK}{{\cal K}}
\newcommand{\sE}{{\cal E}}

\newcommand{\rhor}{\alpha}

\newcommand{\Vol}[2][]{\mathrm{Vol}_{#1} \mleft( #2 \mright) }
\newcommand{\Bocs}[2][]{\mathrm{Box}_{#1} \mleft( #2 \mright) }
\newcommand{\ball}[2][]{#2 {\cal B}_{#1}}
\newcommand{\ballu}[1][]{{\cal B}_{#1}}

\newcommand{\lr}[1]{\mleft( #1 \mright)}
\newcommand{\lrs}[1]{\mleft[ #1 \mright]}
\newcommand{\lrc}[1]{\mleft\{ #1 \mright\}}

\newcommand{\abs}[1]{\mleft| #1 \mright|}
\newcommand{\norm}[1]{\mleft\| #1 \mright\|}

\newcommand{\iV}[2][]{\mathrm{V}_{#1} \mleft( #2 \mright)}
\newcommand{\tiV}[2][]{\widetilde{\mathrm{V}}_{#1} \mleft( #2 \mright)}
\newcommand{\rmax}{r}

\renewcommand{\Re}[1]{\text{Re} \mleft( #1 \mright)}
\renewcommand{\Im}[1]{\text{Im} \mleft( #1 \mright)}

\newenvironment{bsmallmatrix}
  {\mleft[\begin{smallmatrix}}
  {\end{smallmatrix}\mright]}
  
\newcommand{\N}{\eurom{N}}
\newcommand{\A}{\eurom{A}}
\newcommand{\M}{\eurom{M}}
\newcommand{\Pj}{\eurom{P}}

\newcommand{\Le}{l}
\newcommand{\U}{u}

\newcommand{\Urv}{U}
\newcommand{\Lrv}{L}

\begin{document}

\title{A Sphere Packing Bound for Vector Gaussian Fading Channels under Peak Amplitude Constraints}

\author{%
Antonino~Favano\,\orcidlink{0000-0001-7445-7634},~\IEEEmembership{Student Member,~IEEE},
Marco~Ferrari\,\orcidlink{0000-0001-6063-1910},~\IEEEmembership{Member,~IEEE},
Maurizio~Magarini\,\orcidlink{0000-0001-9288-0452},~\IEEEmembership{Member,~IEEE}, and
Luca~Barletta\,\orcidlink{0000-0003-4052-2092},~\IEEEmembership{Member,~IEEE}.
        \thanks{Part of this work was presented at the 2020 IEEE Information
        Theory Workshop~\cite{OurITW20}.}
        \thanks{Antonino Favano is with the Dipartimento di Elettronica,
		Informazione e Bioingegneria, Politecnico di Milano, 20133 Milan, Italy, and also with the Istituto di Elettronica e di Ingegneria dell’Informazione e delle Telecomunicazioni, Consiglio Nazionale delle Ricerche, 20133 Milan, Italy (\mbox{e-mail: }\mbox{\href{mailto:antonino.favano@polimi.it}{\ul{antonino.favano@polimi.it}}}).}
		\thanks{Marco Ferrari is with the Istituto di Elettronica e di Ingegneria dell’Informazione e delle Telecomunicazioni, Consiglio Nazionale delle Ricerche, 20133 Milan, Italy (\mbox{e-mail: }\mbox{\href{mailto:marco.ferrari@ieiit.cnr.it}{\ul{marco.ferrari@ieiit.cnr.it}}}).}
		\thanks{Maurizio Magarini and Luca Barletta are with the Dipartimento di Elettronica,
		Informazione e Bioingegneria, Politecnico di Milano, 20133 Milan, Italy (\mbox{e-mail: }\mbox{\href{mailto:maurizio.magarini@polimi.it}{\ul{maurizio.magarini@polimi.it}}; \href{mailto:luca.barletta@polimi.it}{\ul{luca.barletta@polimi.it}}}).}%
		}%
\maketitle
\begin{abstract}
An upper bound on the capacity of multiple-input multiple-output (MIMO) Gaussian fading channels is derived under peak amplitude constraints. The upper bound is obtained borrowing concepts from convex geometry and it extends to MIMO channels notable results from the geometric analysis on the capacity of scalar Gaussian channels. Relying on a sphere packing argument and on the renowned Steiner's formula, the proposed upper bound depends on the intrinsic volumes of the constraint region, \emph{i.e.}, functionals defining a measure of the geometric features of a convex body. The tightness of the bound is investigated at high signal-to-noise ratio (SNR) for any arbitrary convex amplitude constraint region, for any channel matrix realization, and any dimension of the MIMO system. In addition, two variants of the upper bound are proposed: one is useful to ensure the feasibility in the evaluation of the bound and the other to improve the bound's performance in the low SNR regime. Finally, the upper bound is specialized for two practical transmitter configurations, either employing a single power amplifier for all transmitting antennas or a power amplifier for each antenna.
\end{abstract} 
\section{Introduction}
The capacity of additive white Gaussian noise (AWGN) channels subject to average power constraints is derived by Shannon in~\cite{Shannon1948} and it is perhaps the most renowned accomplishment in information theory. Its extension to multi-antenna systems is another celebrated result and it is presented in~\cite{Telatar1999} by Telatar. Overall, Gaussian channels under average power constraints have been studied extensively. 

On the contrary, the capacity of AWGN channels under peak power or peak amplitude constraints is still an ongoing research topic. Moreover, the information capacity of amplitude-constrained channels is of great interest because of its practical applications, ranging from microwave wireless to free-space optical communications. For more details on the latter, see~\cite{optfreespace1,optfreespace2,optfreespace3,optfreespace4}. In this work, we will mainly focus on the channel capacity of wireless systems.

The ever-growing requirements in data rates and the ubiquitous presence of wireless devices have made energy efficiency one of the fundamental features in the design of a wireless communication system. In this regard, one of the main components affecting the efficiency of the whole system is the power amplifier. Its nonlinear characteristic imposes to limit the peak power of the signals it receives in input. To properly exploit the available resources of a wireless channel, it is fundamental to establish a realistic information theoretic framework, providing an accurate estimate of attainable data rates. By imposing peak power or peak amplitude constraints on the channel input, one can accurately represent the limitations induced by power amplifiers. 

The first significant result in the evaluation of the channel capacity under peak amplitude constraints is presented in~\cite{Smith} by Smith. He proves that the capacity-achieving distribution of an amplitude-constrained scalar Gaussian channel is discrete and comprises a finite number of mass points. Similar results on the discreteness of the capacity-achieving distribution are derived in~\cite{Shamai,Tchamkerten2004,Chan2005,Mamandipoor2014}. In~\cite{McKellips2004}, McKellips presents a simple and tight upper bound on the capacity of scalar amplitude-constrained Gaussian channels. The authors of~\cite{Rassouli} and~\cite{thangaraj2017capacity} derive capacity bounds for multiple-input multiple-output (MIMO) systems with identity channel matrix and subject to a peak amplitude constraint that limits the norm of the input vector. In~\cite{ourISIT2021} for the same MIMO systems and constraint, we provide further insights on the capacity-achieving input distribution and an iterative procedure to numerically evaluate an arbitrarily accurate estimate of the channel capacity. As for systems characterized by nonidentity channel matrices, the authors of~\cite{ElMoslimany2016} investigate the capacity of $2 \times 2$ MIMO systems for rectangular input constraint regions. Furthermore, in~\cite{Dytso} the capacity bounds are further generalized for \mbox{$n$-dimensional} MIMO fading channels subject to any arbitrary peak amplitude constraint. Finally, in~\cite{ourISIT2020} we refine the results presented in~\cite{Dytso} for two particular constraints of practical interest.

\subsection*{Contribution}

In this paper, we provide an asymptotically tight upper bound on the capacity of MIMO AWGN fading channels subject to peak amplitude constraints, that is based on a sphere packing argument and that greatly improves upon the existing literature. We derive the sphere packing (SP) upper bound by extending the results in~\cite{rhosigma} to MIMO systems. We prove that the gap between the derived upper bound and the best lower bound from the literature~\cite{ourISIT2020}, is asymptotically tight. Indeed, as the signal-to-noise ratio (SNR) goes to infinity, the bounds' gap vanishes for any channel matrix, any dimension of the MIMO system, and any convex constraint region. Moreover, we introduce two variants of the SP upper bound. One is useful to always guarantee a feasible evaluation of the bound and the other is able to improve the performance at the low SNR. We also specialize the SP bound for two practical transmitter configurations. One employs a single amplifier for all transmitting antennas, which determines a constraint on the norm of the input vector. We refer to it as \emph{total amplitude} (TA) constraint. The second transmitter configuration employs multiple power amplifiers, one per transmitting antenna. It induces a peak amplitude constraint on each entry of the input vector, which we define as \emph{\mbox{per-antenna}} (PA) constraint.

\subsection*{Structure of the Paper}

In Section~\ref{S:prelim}, we introduce some useful mathematical tools needed throughout the rest of the paper; in Section~\ref{S:sysmod} we define the system model; in Section~\ref{S:SoA} we outline previous results present in the current literature; and in Section~\ref{S:SPbound} we describe the SP upper bound and its variants. In Section~\ref{S:SpecA}, we specialize the SP bound for the TA constraint and for the PA constraint. Furthermore, we evaluate the tightness of the resulting capacity bounds and compare it with the main results from previous works. Section~\ref{S:conclusion} concludes the paper.

\subsection*{Notation}
We use bold letters for vectors ($\xv$), uppercase letters for random variables ($X$), calligraphic uppercase letters for subsets of vector spaces ($\sX$), and uppercase sans serif letters for matrices ($\Hv$). We denote by $\norm{\xv}$ the Euclidean norm of the vector $\xv$, by $\Hv^T$ the transposed of a matrix $\Hv$, and by $\det\lr{\Hv}$ its determinant. Furthermore, ${\cal CN}(\boldsymbol{\mu},\mathsf{\Sigma})$ indicates a complex multivariate Gaussian distribution with mean vector $\boldsymbol{\mu}$ and covariance matrix $\mathsf{\Sigma}$. We represent the $n \times 1$ vector of zeros by $\textbf{\textsf{0}}_n$ and the $n \times n$ identity matrix by $\Iv_n$. We denote by $\ballu[n]\triangleq \lrc{\xv : \norm{\xv} \leq 1}$ the \mbox{$n$-dimensional} unit ball in $\mathbb{R}^{n}$ centered in $\textbf{\textsf{0}}_{n}$ and by $\ball[n]{\delta}\triangleq \lrc{\xv : \norm{ \xv } \leq \delta}$ the \mbox{$n$-dimensional} ball of radius $\delta$ centered in $\textbf{\textsf{0}}_{n}$. We define the \mbox{$n$-dimensional} box of sides $\bm{d}$ as $\Bocs[n]{\bm{d}} \triangleq \{ \xv :  \abs{x_k} \leq d_k/2 , \ k=1,\dots,n \}$ and, with a slight abuse of notation, we use $\Bocs[n]{d}$ whenever $d_k = d, \ \forall k$. Finally, we define $\Hv \sX \triangleq \{ \yv : \yv = \Hv \xv, \, \xv \in \sX  \} $, we denote the $\M$-times Cartesian product of $\Hv \sX$ with itself by $[\Hv \sX]^{\times \M}$, and the \mbox{$n$-dimensional} volume of the set $\sX$ by $\Vol[n]{\sX}$.

\section{Preliminaries} \label{S:prelim}

Given two subsets $\sK$ and ${\cal R}$ of a vector space, the Minkowski sum is denoted by the operator~$\oplus$ and it gives the set obtained by adding each vector in $\sK$ to each vector in~$\cal R$ as
\begin{align}
  \sK \oplus {\cal R} \triangleq \lrc{ {\bf k} + {\bf r} \mid {\bf k} \in \sK , \ {\bf r} \in {\cal R} }.   
\end{align}
Moreover, if $\sK$ is a convex body in $\mathbb{R}^n$, we denote by $\iV[j]{\sK}$ the $j$th intrinsic volume of $\sK$, with $j=0,\dots,n$. Intrinsic volumes are nonnegative, homogeneous, and monotonic functionals and represent a fundamental measure of content for a convex body~\cite{lotz2020concentration}. The authors of~\cite{lotz2020concentration} also provide an intuitive, yet technical, definition as follows. Let $\mathsf{P}_j$ be an $n \times n$ orthogonal projection matrix, projecting onto a fixed $j$-dimensional subspace of $\mathbb{R}^n$. Furthermore, let $\mathsf{Q}$ be an $n \times n$ random rotation matrix drawn uniformly from the Haar measure\footnote{For a rigorous definition of the Haar measure, see~\cite{conway2019course}.} on the compact, homogeneous group of $n \times n$ orthogonal matrices with determinant one. Then, the intrinsic volumes of a convex body $\sK$ are defined as
\begin{align} 
\iV[j]{\sK} \triangleq \binom{n}{j} \frac{\kappa_n}{\kappa_j\kappa_{n-j}} \expect{\Vol[j]{\mathsf{P}_j\mathsf{Q}\sK}}, \quad j=0,\dots,n, \label{eq:iVdef}
\end{align}
where the expectation is taken with respect to the random rotation matrix $\mathsf{Q}$ and where $\kappa_i \triangleq \pi^{\frac{i}{2}}/\Gammaf{\frac{i}{2}+1}$ is the volume of the \mbox{$i$-dimensional} unit ball. We remark that from~\eqref{eq:iVdef}, it is intuitive to see that the $n$th intrinsic volume of $\sK$ coincides with its volume $\Vol[n]{\sK}$. They are characterized by the following property
\begin{align} \label{eq:homogeneous}
    \iV[j]{\A \sK} = \A^j \iV[j]{\sK}, \ \A \geq 0, \ \forall j.
\end{align}
Given a set $\sK$ such that $\sK \supset \sR$, it holds
\begin{align} \label{eq:monotonicity}
    \iV[j]{\sK} \geq \iV[j]{\sR}, \ \forall j.
\end{align}
Thanks to the Steiner's formula~\mbox{\cite[Theorem~4]{rhosigma}} we can evaluate the \mbox{$n$-dimensional} volume of the Minkowski sum of a convex set $\sK$ with a ball or radius $\delta \ge 0$ as follows
\begin{align} \label{eq:Steiner_formula}
\Vol[n]{ \sK \oplus \ball[n]{\delta}} = \sum_{j=0}^{n} \iV[j]{\sK} \Vol[n-j]{\ball[n-j]{\delta}}.
\end{align}
Since~\eqref{eq:Steiner_formula} is a convolution, it is useful to introduce the (logarithmic) generating function of the intrinsic volumes of $\sK$ as~\mbox{\cite[Theorem~8]{rhosigma}}
\begin{align}
&\G_\sK(t) = \log  \lr{ \sum_{j=0}^n \iV[j]{\sK}e^{jt}  }.
\end{align}
As shown in~\cite{lotz2020concentration}, an important property of these generating functions is that, given two sets $\sK$ and ${\cal R}$, it holds
\begin{align} \label{eq:propGenF}
   \G_{\sK\times {\cal R}}(t) = \G_{\sK}(t) + \G_{\cal R}(t), \ \forall t \in \mathbb{R}. 
\end{align}
Finally, given a function $\f:\mathcal{T}\to\mathbb{R}\cup\lrc{-\infty,+\infty}$, from~\cite{rockafellar1997convex} we define the convex conjugate of $\f$ as
\begin{align} \label{eq:convexconj}
    \f^*(t^*) \triangleq \sup_{t \in \mathcal{T}} \lrc{t\cdot t^* - \f(t)},
\end{align}
where $\f^*:\mathcal{T}^*\to\mathbb{R}\cup\lrc{-\infty,+\infty}$, with $\mathcal{T}^*$ being the dual space to $\mathcal{T}$.

\section{System Model} \label{S:sysmod}
Let us consider an $\N \times \N$ complex MIMO system with input-output relationship given by
\begin{align} \label{eq:model}
\tYv = \widetilde{ \Hv } \cdot \tXv + \tZv.
\end{align}
The input vector $\tXv$ is such that $\tXv \in \widetilde{\sX}$ with $\widetilde{\sX}$ being a convex constraint region, $\tZv \sim {\cal CN}(\textbf{\textsf{0}}_{\N},2 \varn \Iv_{\N})$ is the noise vector, and $\widetilde{ \Hv } $ is any full rank channel fading matrix. We assume $\widetilde{ \Hv } $ to be constant throughout the channel uses and known both at the transmitter and at the receiver. Let us vectorize the system in~\eqref{eq:model} in its real and imaginary components. We obtain the equivalent model
\begin{align} \label{eq:eqmodel}
\Yv = \Hv \cdot \Xv + \Zv,
\end{align}
with $\Hv = \text{Re}\{ \widetilde{ \Hv } \} \otimes \Iv_{2} + \text{Im}\{ \widetilde{ \Hv } \} \otimes \begin{bsmallmatrix}0 & -1\\1 & \ 0\end{bsmallmatrix}$, where the operator~$\otimes$ is the Kronecker product, $\Yv$ is a $2\N \times 1$ vector defined as $\Yv = [\text{Re}\{\widetilde{Y}_1\},\text{Im}\{\widetilde{Y}_1\},\dots,\text{Re}\{\widetilde{Y}_\N\},\text{Im}\{\widetilde{Y}_\N\}]^T$, and analogously for $\Xv$ and $\Zv$. 

We define the MIMO channel capacity as
\begin{align} \label{eq:capacity}
    \C\lr{\sX,\Hv,\varn} &\triangleq \max_{F_\Xv: \: \text{supp}(F_\Xv)\subseteq {\sX}} \mi{\Xv}{\Yv}, 
\end{align}
where $F_\Xv$ is the input distribution law and $\sX$ is the equivalent input constraint region derived from $\widetilde{ \sX }$. Since we investigate the capacity of amplitude-constrained channels, the peak power of the input signal is a more relevant parameter than its average power. Therefore, it is convenient to define the SNR as the ratio between the peak power of the input signal and the trace of the noise covariance matrix. Then, the SNR depends on the specific constraint region $\sX$ and it is $\text{SNR} \triangleq (r_\text{max} (\sX) )^2 / (2 \N\varn )$, with $r_\text{max} (\sX) \triangleq \sup_{\xv \in \sX}\lrc{ \norm{ \xv } }$.

\section{Previously Proposed Bounds} \label{S:SoA}

In~\cite{Dytso}, the authors provide capacity upper and lower bounds for AWGN MIMO systems under an arbitrary peak amplitude constraint region $\sX$ and for any channel matrix, known at both the transmitter and the receiver. Their duality upper bounds are derived by considering an enlarged output constraint region $\sD \supset \Hv \sX$. The advantage of using $\sD$ is that it can have a simpler structure than $\Hv \sX$, making the derivation of an upper bound feasible. Specifically, they consider $\sD$ to be either a ball or a box. Let $\sD_1$ be the $2\N$-dimensional ball of radius $d_1 = r_\text{max} \lr{ \Hv \sX}$ and $\sD_2 = \Bocs[2\N]{\mathbf{d}_2}$ the smallest box containing $\Hv \sX$. Then, in~\mbox{\cite[Theorem~10]{Dytso}} the authors define their duality upper bounds as follows
\begin{align} \label{eq:DUBball}
    \C \leq \overline{\C}_{\text{D,1}} \triangleq \log \lr{c_{2\N}(d_1) + \frac{\Vol[2\N]{\sD_1}}{\lr{2 \pi e \varn}^{\N}} },
\end{align}
where $c_{2\N}(d_1) = \sum_{j=0}^{2\N-1}\binom{2\N-1}{j}\frac{\Gammaf{\N-j/2}}{2^{j/2}\Gammaf{\N}} \lr{d_1 / \sigma_z}^j$ and
\begin{align} \label{eq:DUBbox}
    \C \leq \overline{\C}_{\text{D,2}} \triangleq \sum_{j=1}^{2\N} \log \lr{1+\frac{d_{2,j}}{\sqrt{2 \pi e \varn}}},
\end{align}
where $d_{2,j}$ is the $j$th component of $\mathbf{d}_2$. The main disadvantage of these bounds is that the more $\sD$ differs from $\Hv \sX$, the less accurate the bounds become. In~\cite{ourISIT2020}, by extending the McKellips-Type upper bound of~\cite{thangaraj2017capacity} to MIMO systems affected by fading, we improve the results of~\cite{Dytso} for specific cases of the PA constraint and for the TA constraints. Intuitively, we achieve better results than those in~\cite{Dytso} by considering an upper bound that depends on a smaller constraint region $\sS$ such that $\sD \supset \sS \supset \Hv \sX$. Although the resulting asymptotic capacity gap in~\cite{ourISIT2020} is smaller compared to the past literature, it can be far from zero. Moreover, it widens as $\N$ grows larger and the derived upper bounds are valid just for the mentioned specific cases. 

In the next section, we define an upper bound that improves upon those of~\cite{ourISIT2020} and that is also far more general, as it is for the duality upper bounds in~\cite{Dytso}. We prove that our upper bound provides an asymptotic gap equal to zero for any convex constraint region, any channel matrix $\Hv$, and any MIMO dimension $\N$. 

\section{Sphere Packing Upper Bound}
\label{S:SPbound}
In~\cite{rhosigma}, the authors investigate the capacity of AWGN scalar channels under average and peak power constraints. They provide an upper bound based on an SP argument by using a fundamental result from convex geometry, \emph{i.e.}, the Steiner's formula in~\eqref{eq:Steiner_formula}. We extend their upper bound to MIMO fading channels. By considering an arbitrarily large number, $\M$, of independent channel uses, for $\M \to \infty$, the SP bound is 
\begin{align} \label{eq:SPbound}
\C \leq \overline{\C}_\text{SP} &\triangleq \limsup_{\M \to \infty}\frac{1}{\M} \log{\frac{\Vol[n]{ [\Hv \sX]^{\times \M} \oplus \ball[n]{\delta}}}{\Vol[n]{\ball[n]{\delta}}} } \\
&= \limsup_{\M \to \infty}\frac{1}{\M}\log{\Vol[n]{ [\Hv \sX]^{\times \M} \oplus \ball[n]{\delta}}} \label{eq:10} \nonumber \\ & \quad - \lim_{\M \to \infty}\frac{1}{\M} \log{  {\Vol[n]{\ball[n]{\delta}}} } \\
&= \Ell(\varn) - \N\log \lr{ 2 \pi e \varn  }, \label{eq:finalSPbound}
\end{align}
where $n=2 \N \M$ and $\delta=\sqrt{n \varn}$. Let us focus on the evaluation of the term $\Ell(\varn)$ in~\eqref{eq:finalSPbound}. To deal with the convolution in~\eqref{eq:10} involving the output signal space $\sK = \lrs{\Hv \sX}^{\times \M}$, we define the limiting normalized generating function of the intrinsic volumes of $\sK$, $\func$, as
\begin{align}
\func &\triangleq \lim_{\M \to \infty} \frac{1}{2 \N \M} \G_\sK(t) \\
&= \lim_{\M \to \infty} \frac{1}{2 \N \M} \G_{\lrs{\Hv \sX}^{\times \M}}(t) \\
&\stackrel{\eqref{eq:propGenF}}{=} \lim_{\M \to \infty} \frac{\M}{2 \N \M} \G_{\Hv \sX}(t) \label{eq:stackrel}\\
&= \frac{1}{2 \N } \log  \bigg( \sum_{j=0}^{2 \N } \iV[j]{\Hv \sX}e^{jt}  \bigg), \label{eq:16}
\end{align}
where $\stackrel{\eqref{eq:propGenF}}{=}$ indicates that~\eqref{eq:stackrel} holds thanks to~\eqref{eq:propGenF}. Note that the limit exists because, being $\Hv \sX$ a finite set, its intrinsic volumes $\iV[j]{\Hv \sX}$ exist and are finite as well. By following the steps in~\mbox{\cite[Lemma~14]{rhosigma}}, let us define the sequence of measures $\tiV[\theta]{\sK}\triangleq\tiV[j/n]{\sK}=\iV[j]{\sK}$ with $ j = 0,\dots, n $ and $\theta \in [0,1]$. Furthermore, let $\convf$ be the convex conjugate of $\func$, as defined in~\eqref{eq:convexconj}. We have
\begin{align}
    \lim_{\M \to \infty} \frac{1}{\M} \log \tiV[\theta]{\sK} &= \lim_{\frac{n}{2 \N } \to \infty} \frac{2 \N }{n} \log \tiV[\theta]{\sK} \\
    &= 2\N \lim_{n \to \infty} \frac{1}{n} \log \tiV[\theta]{\sK} \\
    &= 2\N \sup_\theta \lrc{ -\convf }, \label{eq:GEtheo}
\end{align}
where~\eqref{eq:GEtheo} holds by~\mbox{\cite[Lemma~14]{rhosigma}} and~\mbox{\cite[Lemma~15]{rhosigma}}, which build on the \mbox{Gärtner-Ellis} large deviations theorem. The application of these lemmas requires that the limit $\f(t)$ exists for any $t \in \mathbb{R}$ and that $\f(0)< \infty$. By~\eqref{eq:16}, it is clear that, in our case, these requirements are always satisfied. 

We now have all the necessary elements to evaluate $\Ell(\varn)$. By applying~\mbox{\cite[Theorem~18]{rhosigma}}, for $\varn>0$ it holds that
\begin{align}
&\Ell(\varn) = \limsup_{\M \to \infty}\frac{1}{\M}\log{\Vol[n]{ [\Hv \sX]^{\times \M} \oplus \ball[n]{\delta}}} \\ 
& \ \stackrel{\eqref{eq:GEtheo}}{=} \sup_{\theta \in [0,1]} \lrc{ -2\N \convf + (1-\theta)\N \log \frac{2\pi e\varn}{1-\theta} } \\ 
& \ \stackrel{\eqref{eq:convexconj}}{=} \sup_{\theta \in [0,1]} \bigg\{ -2 \N \sup_t \lrc{ \theta t - \func} + (1-\theta)\N  \log \frac{2\pi e\varn}{1-\theta} \bigg\} \\
& \ \stackrel{\eqref{eq:16}}{=} \sup_{\theta \in [0,1]} \bigg\{ -2 \N \sup_t \bigg\{ \theta t - \frac{1}{2 \N } \log \bigg(  \sum_{j=0}^{2 \N } \iV[j]{\Hv \sX}e^{jt} \bigg) \bigg\} + (1-\theta)\N  \log \frac{2\pi e\varn}{1-\theta} \bigg\}. \label{eq:finalL}
\end{align}
Finally, by plugging~\eqref{eq:finalL} into~\eqref{eq:finalSPbound} we obtain the final bound. 

\subsection{Asymptotic Gap}

Let us now evaluate the asymptotic gap between the SP bound, denoted by $\overline{\C}_\text{SP}$, and the entropy power inequality~(EPI) lower bound, which is derived in~\cite{Dytso} as
\begin{align} \label{eq:Cepi}
    \underline{\C}_\text{EPI} = \N \log \lr{1+\frac{\lr{\Vol[2 \N ]{\Hv \sX}}^{\frac{1}{\N}}}{2 \pi e \varn}}.
\end{align}
By~\mbox{\cite[Lemma~18]{rhosigma}} we have that, for $\varn \to 0$, the maximizing $\theta$ in~\eqref{eq:finalL} is $1$ and therefore it holds that
\begin{align} \label{eq:L0}
\lim_{\varn \to 0} \Ell(\varn)
&= -2\N \f^*(1) \\
&= -2\N \sup_t \lrc{ t - \frac{1}{2 \N } \log \lr{  \sum_{j=0}^{2 \N } \iV[j]{\Hv \sX}e^{jt} } } \\
&= \inf_t \lrc{ \log \lr{  \sum_{j=0}^{2 \N } \iV[j]{\Hv \sX}e^{(j-2\N)t} } } \\
&=  \log \lr{ \inf_t \lrc{ \sum_{j=0}^{2 \N } \iV[j]{\Hv \sX}e^{(j-2\N)t} } } \\
&=  \log \lr{ \iV[2\N]{\Hv \sX} + \inf_t \lrc{ \sum_{j=0}^{2 \N -1} \iV[j]{\Hv \sX}e^{(j-2\N)t} } } \label{eq:pregapVOL} \\ 
&=  \log \lr{\Vol[2 \N]{\Hv \sX}}, \label{eq:gapVOL}
\end{align}
where~\eqref{eq:gapVOL} holds because the argument of the infimum in~\eqref{eq:pregapVOL} is a sum of exponentials scaled by nonnegative coefficients, therefore the sum is minimized in $t \to -\infty$ and the infimum is zero.
The gap at high SNR results in
\begin{align}
    g_\text{SP} &\triangleq \lim_{\varn \to 0} \overline{\C}_\text{SP} - \underline{\C}_\text{EPI} \\
    &= \lim_{\varn \to 0} \Ell(\varn) - \N \log \lr{\lr{\Vol[2 \N ]{\Hv \sX}}^{\frac{1}{\N}}} = 0. \label{eq:0gap}
\end{align}
Therefore, we proved that the SP upper bound is asymptotically tight at high SNR for any dimension $\N$ of the MIMO system, any channel matrix $\Hv$,
and any convex constraint region $\sX$.

\subsection{Generalized Sphere Packing}
\label{S:GA}

A practical issue in the evaluation of the SP upper bound is that it is not always trivial to derive the intrinsic volumes of a given region. Furthermore, even when the intrinsic volumes of $\sX$ are known, the distortion induced by the channel matrix $\Hv$ can anyway make the evaluation of the intrinsic volumes of $\Hv \sX$ quite complicated, if not unfeasible. Therefore, in this subsection we introduce a further upper bound on the SP approach, that relies just on being able to evaluate $\Vol{\sX}$.

The core idea is somewhat reminiscent of the one used in the upper bounds of~\cite{Dytso}. To evaluate their duality upper bounds, the authors of~\cite{Dytso} replace $\Hv \sX$ with a region $\sD \supset \Hv \sX$. Roughly speaking, introducing the region $\sD$ is similar to replacing all the intrinsic volumes of $\Hv \sX$ with $\iV[j]{\sD} \geq \iV[j]{\Hv \sX}$. Instead, in our generalized sphere packing (G-SP) approach, we are free to upper-bound each $\iV[j]{\Hv \sX}$ independently. Aside from the derived improved flexibility, the G-SP approach is especially useful because it allows us to keep the volume of $\Hv \sX$ unaltered, which lets us retain the desirable asymptotic properties of the SP upper bound. We remark that $\iV[0]{\sK} = 1$ for any $\sK$ and that $\iV[2 \N]{\Hv \sX} = \det\lr{\Hv} \Vol[2 \N ]{\sX}$ can be typically evaluated. Instead, for the $\iV[j]{\Hv \sX}$'s, with $j = 1,\dots, 2\N-1$, we can always define an upper bound by choosing, for each $j$, an appropriate region $\sS_j \supset \Hv \sX$, which by~\eqref{eq:monotonicity} guarantees that $\iV[j]{\sS_j} \geq \iV[j]{\Hv \sX}$. Therefore, we have
\begin{align} \label{eq:geniVUB}
\iV[j]{\Hv \sX}\leq
    \begin{cases}
    1, &j=0 \\
    \iV[j]{\sS_j}, &j=1,\dots,2 \N -1 \\
    \det\lr{\Hv} \Vol[2 \N ]{\sX}, &j= 2 \N.
    \end{cases}
\end{align}
Finally, we define the G-SP upper bound as
\begin{align} \label{eq:G-SP}
    \overline{\C}_\text{G-SP} \triangleq \Ell_{\text{G}}(\varn) - \N\log \lr{ 2 \pi e \varn  },
\end{align}
where $\Ell_{\text{G}}(\varn)$ is simply derived by plugging the right-hand side of~\eqref{eq:geniVUB} into~\eqref{eq:finalL}. Notice that, since the logarithm is a monotonic function and the intrinsic volumes in~\eqref{eq:finalL} are nonnegative coefficients in a sum of exponentials, it holds $\Ell_{\text{G}}(\varn) \geq \Ell(\varn)$ and therefore $\overline{\C}_\text{G-SP} \geq \overline{\C}_\text{SP}$.

As for the asymptotic gap behavior of the G-SP bound, we have
\begin{align}
    g_\text{G-SP} &\triangleq \lim_{\varn \to 0} \overline{\C}_\text{G-SP} - \underline{\C}_\text{EPI} = g_\text{SP} = 0.
\end{align}
Indeed, as $\varn$ goes to zero we get the same result of~\eqref{eq:gapVOL} for $\Ell_{\text{G}}(\varn)$ as well. Then, the G-SP upper bound asymptotically depends uniquely on the $2 \N $th intrinsic volume, which by~\eqref{eq:geniVUB} is still $\iV[2\N]{\Hv \sX}$ and therefore determines the same vanishing gap of~\eqref{eq:0gap}.

\subsection{Piecewise Sphere Packing} 
\label{sss:P-SP}

\begin{figure}[t]
	\centering
	\includegraphics[width=0.45\linewidth]{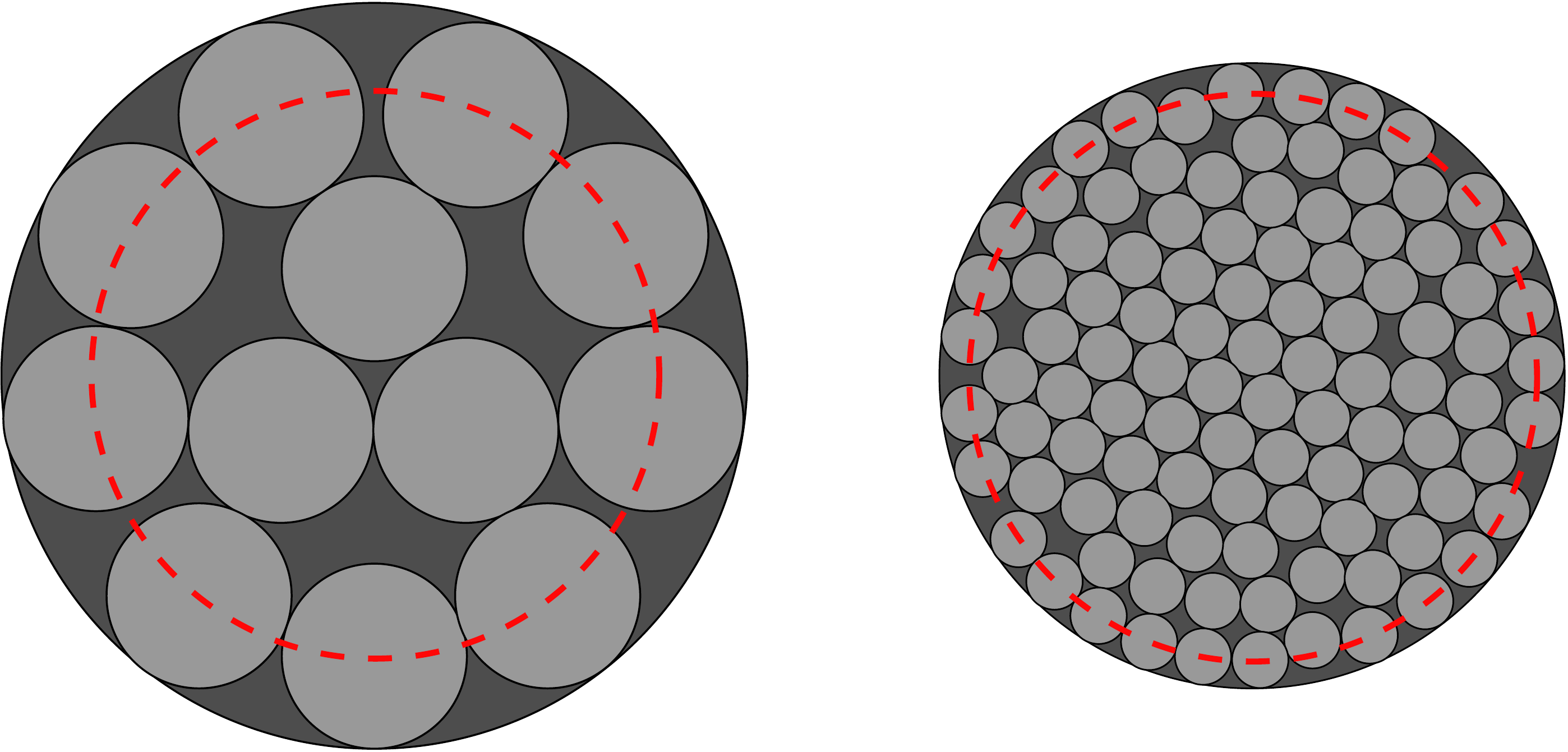}
	\caption{Two sphere packing examples under a peak amplitude constraint and different SNR values. For both cases, the red dashed line is the border of $\sK = \ball[2]{}$, in dark gray the result of the Minkowski sum between $\sK$ and the noise ball for each specific SNR. On the left, the light gray noise balls are translated replicas of $\ball[2]{\delta_1}$, while on the right of $\ball[2]{\delta_2}$ with $\delta_1>\delta_2$.}
	\label{fig:SPminksum}
\end{figure}

Although the SP and G-SP upper bounds are asymptotically tight, they can be loose at low SNR. Indeed, since the SP bound is based on geometric arguments, its reliability depends on how accurately the Minkowski sum approximates the true channel output region. Intuitively, the Minkowski sum in~\eqref{eq:SPbound} is obtained by taking the union of an infinite number of noise balls $\ball[n]{\delta}$, after centering them in each point of $\sK = [\Hv \sX]^{\times \M}$. Conversely, a true sphere packing problem would take the union of nonoverlapping replicas of $\ball[n]{\delta}$, packed in $[\Hv \sX]^{\times \M} \oplus \ball[n]{\delta}$. Therefore, it is intuitive that, as the noise balls get smaller, the Minkowski sum becomes a better approximation of the output signal space $[\Hv {\cal X}]^{\times \M}$, as shown in Fig.~\ref{fig:SPminksum}. At the same time, Fig.~\ref{fig:SPminksum} also shows that the Minkowski sum can be far from an ideal approximation at low SNR.
We now derive a family of piecewise sphere packing (P-SP) upper bounds that is able to improve upon the standard SP bound in~\eqref{eq:finalSPbound} in the mentioned SNR range.

Let us separate the MIMO channel into two independent subchannels. Then, we can apply the SP upper bound on one subchannel and the Gaussian maximum-entropy bound on the other. Given the MIMO channel capacity
\begin{align}
    \C 
    &= \max_{F_\Xv : \: \Xv \in \sX} \lrc{ \ent{\Yv} } - \ent{\Zv},
\end{align}
an upper bound on the first term on the right-hand side is given by
\begin{align}
   \max_{F_\Xv : \: \Xv \in \sX} \lrc{ \ent{\Yv} } &= \max_{F_\Xv : \: \Xv \in \sX} \lrc{ \ent{\Yv_\Urv,\Yv_\Lrv} } \\
   &\leq \max_{F_\Xv : \: \Xv \in \sX} \lrc{ \ent{\Yv_\Urv} + \ent{\Yv_\Lrv} }, \label{eq:hYhYl}
\end{align}
where $\Yv = \lrs{ \Yv_\Urv , \Yv_\Lrv }^T$, with $\Yv_\Urv = \lrs{Y_1,\dots,Y_{\U}}^T \in \mathbb{R}^{\U}$, $\Yv_\Lrv = \lrs{Y_{\U+1},\dots,Y_{\U+\Le}}^T \in \mathbb{R}^{\Le}$, and $\U+\Le = 2 \N $. We want to treat independently the contributions of $\ent{\Yv_\Urv}$ and $\ent{\Yv_\Lrv}$, to upper-bound them with the two mentioned techniques. First, we notice that given the singular value decomposition of $\Hv = \Uv \Dv \Vv^T$, it holds
\begin{align}
    \iV[j]{\Hv \sX} = \iV[j]{\Dv \sX}, \quad \forall j.
\end{align}
We denote the diagonal elements of $\Dv$ as $\lambda_1, \dots, \lambda_{2 \N }$ and we assume that $\lambda_1 \geq \lambda_2 \geq \dots \geq \lambda_{2 \N }$, since it is always possible to rearrange the MIMO system in such a way that this condition is satisfied. Then, the subsystem to which $\Yv_\Urv$ belongs, perceives larger singular values, and therefore higher SNR. On the contrary, $\Yv_\Lrv$ refers to the subsystem affected by a lower SNR due to smaller singular values. Then, we expect the SP upper bound to be more accurate on $\ent{\Yv_\Urv}$, where the transmitted signal is stronger compared to the noise level, while we expect it to be less precise on $\ent{\Yv_\Lrv}$. Since in the subsystem of $\Yv_\Lrv$ the Gaussian noise is dominant, a tighter upper bound on $\ent{\Yv_\Lrv}$ can be provided by the differential entropy of a normally distributed vector $\overline{\Yv}_\Lrv \sim {\cal N} \lr{\textbf{\textsf{0}}_{\Le} , \mathsf{\Sigma}_\Lrv }$, with $\mathsf{\Sigma}_\Lrv = \Dv_\Lrv \expect{\Xv_\Lrv\Xv_\Lrv^T} + \varn\Iv_\Le$ and with $\Dv_\Lrv$ being the $\Le \times \Le$ submatrix of $\Dv$ with diagonal elements $\lambda_{\U+1},\dots,\lambda_{2 \N }$. Furthermore, to separate the capacity contributions of the two subsystems, we need to reformulate the input constraint in such a way that it can be separated as well. Let us assume $\sX$ to be a ball and let us define its radius as $\rmax = r_{\text{max}}\lr{\sX}$. Then, we have
\begin{align}
     \norm{ \Xv }^2 =  \norm{ \Xv_\Urv }^2 +  \norm{ \Xv_\Lrv }^2 \leq \rmax^2.
\end{align}
We reinterpret $\rmax^2$ as $\rmax^2 (1-\rhor)^2 + \rmax^2 \rhor^2$ with $\rhor \in [0,1]$. Therefore, the constraint $ \norm{ \Xv } \leq \rmax$ becomes equivalent to
\begin{align}
\begin{cases}
 \norm{ \Xv_\Urv } \leq \rmax \lr{1-\rhor} \\
 \norm{ \Xv_\Lrv } \leq \rmax \rhor.
\end{cases}
\end{align}
By plugging this equivalent constraint into~\eqref{eq:hYhYl}, we obtain
\begin{align}
&\max_{F_\Xv : \: \Xv \in \sX} \lrc{ \ent{\Yv_\Urv} + \ent{\Yv_\Lrv} } \\
& \quad = \max_{\rhor \in [0,1]} \bigg\{ \max_{F_\Xv : {\scriptsize \begin{cases}
\norm{ \Xv_\Urv } \leq \rmax \lr{1-\rhor}  \\
\norm{ \Xv_\Lrv } \leq \rmax \rhor
\end{cases}}} \lrc{ \ent{\Yv_\Urv} + \ent{\Yv_\Lrv} } \bigg\} \\
& \quad = \max_{\rhor \in [0,1]} \bigg\{ \max_{F_{\Xv_\Urv} : \: \norm{ \Xv_\Urv } \leq \rmax \lr{1-\rhor} } \lrc{ \ent{\Yv_\Urv} } + \max_{F_{\Xv_\Lrv} : \: \norm{ \Xv_\Lrv } \leq \rmax \rhor } \lrc{ \ent{\Yv_\Lrv} } \bigg\}.
\end{align}
Then, we can apply the SP upper bound in~\eqref{eq:SPbound} on the subsystem of $\Yv_\Urv$ to get
\begin{align}
   &\max_{F_{\Xv_\Urv} : \: \norm{ \Xv_\Urv } \leq \rmax \lr{1-\rhor} } \lrc{ \ent{\Yv_\Urv} } \leq  \Ell_\Urv(\rhor) \\
   & \quad = \sup_{\theta \in [0,1]} \bigg\{ -\U\sup_t \bigg\{ \theta t - \frac{1}{\U} \log \sum_{j=0}^{\U} \iV[j]{\Dv_\Urv \sX_\Urv}e^{jt} \bigg\} + (1-\theta)\frac{\U}{2}  \log \frac{2\pi e\varn}{1-\theta} \bigg\},
\end{align}
where $\Dv_\Urv$ is the $\U \times \U$ submatrix of $\Dv$ with diagonal elements $\lambda_1, \dots, \lambda_U$ and $\sX_\Urv = \ball[\U]{\rmax(1-\rhor)}$. For the subsystem associated with $\Yv_\Lrv$, the upper bound is given by
\begin{align}
& \max_{F_{\Xv_\Lrv} : \: \norm{ \Xv_\Lrv } \leq \rmax \rhor } \lrc{ \ent{\Yv_\Lrv} } \leq \max_{F_{\Xv_\Lrv} : \: \norm{ \Xv_\Lrv } \leq \rmax \rhor } \lrc{ \ent{\overline{\Yv}_\Lrv} } \\
& \ = \max_{\substack{\expect{  \norm{\Xv_\Lrv}^2} : \\ \norm{ \Xv_\Lrv } \leq \rmax \rhor }} \sum_{k=1}^{\Le} \frac{1}{2} \log \lr{2\pi e \lr{\lambda^2_{u+k}\expect{ \abs{X_{\Lrv,k}}^2}+\varn}} \\
& \ = \max_{\substack{\expect{  \norm{\Xv_\Lrv}^2} : \\ \norm{ \Xv_\Lrv } \leq \rmax \rhor }} \sum_{j=\U+1}^{2 \N } \frac{1}{2} \log \lr{2\pi e \lr{\lambda^2_j\rhor^2\expect{ \abs{X_j}^2}+\varn}} \label{eq:WFmax} \\
& \ = \sum_{j=\U+1}^{2 \N } \frac{1}{2}\log \lr{2\pi e \lr{\lambda^2_j\rhor^2\Pj_j+\varn}},
\end{align}
where $X_{\Lrv,k}$ is the $k$th component of the vector $\Xv_{\Lrv}$ and $\Pj_j$ is the power allocation given by the water-filling algorithm, which maximizes~\eqref{eq:WFmax} for the given constraint $\norm{ \Xv_\Lrv } \leq \rmax \rhor$. Notice that we obtain a valid upper bound for any combination of positive integers $\U$ and $\Le$ with sum equal to $2 \N $, therefore the complete formulation of the P-SP upper bound is
\begin{align} \label{eq:P-SP}
    \C \leq \overline{\C}_\text{P-SP} &= \min_{\substack{\U: \\ \U+\Le = 2 \N }} \max_{\rhor} \Bigg\{ \Ell_\Urv(\rhor) + \sum_{j=\U+1}^{2 \N } \frac{1}{2} \log \lr{2\pi e \lr{\lambda^2_j\rhor^2 \Pj_j+\varn}} \Bigg\} - \N \log(2\pi e \varn).
\end{align}
 Finally, we remark that even when $\sX$ is not a ball we can simply consider the enlarged constraint $\ball[2 \N]{r_\text{max}(\sX)} \supset \sX$ instead of $\sX$ and still provide a valid upper bound.
 
\section{Practical Cases} 
\label{S:SpecA}

We now evaluate the performance of the SP and G-SP upper bound subject to specific constraints, induced by common transmitter configurations. Specifically, let us focus on two mentioned cases of practical interest, namely the \emph{total amplitude} (TA) and the \emph{per-antenna} (PA) constraint.

\subsection{Total Amplitude Constraint}
\label{S:TAbound}

\begin{figure}[t]
    \centering
    \input{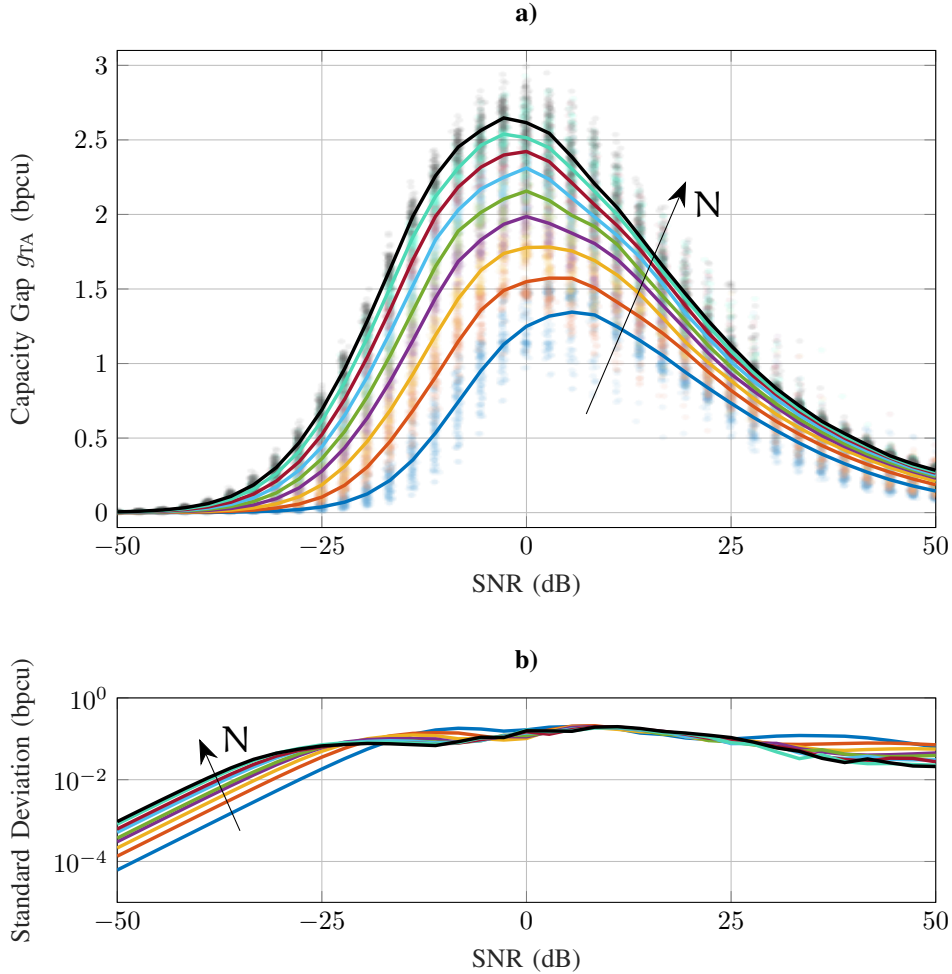}
    \caption{\textbf{a)} Numerical evaluation of the capacity gap $g_\text{TA}$, defined in~\eqref{eq:g_TA}, in bit per channel use (bpcu) versus SNR, for $\N=2,\dots,10$. For each $\N$, the filled circles are the gaps resulting from each random channel realization, while the solid lines show the averaged behavior. \textbf{b)} Standard deviation of $g_\text{TA}$ in bpcu versus SNR, for $\N=2,\dots,10$.}
    \label{fig:MCgapTA}
\end{figure}%
\begin{figure}[t]
    \centering
    \input{Figures/TA_CapacityGap_overN.tex}
    \caption{Numerical evaluation of the average capacity gap per complex dimension in bit per channel use (bpcu) versus SNR, for $\N=2,\dots,10$. The solid lines are $\expect{g_\text{TA}}/\N$, with $g_\text{TA}$ defined in~\eqref{eq:g_TA}. The dashed lines are $\expect{g_\text{D,TA}}/\N$, with $g_\text{D,TA}$ defined in~\eqref{eq:gD_TA}.}
    \label{fig:MCgap/N_TA}
\end{figure}%
\begin{figure}[t]
    \centering
    \input{Figures/TA_GAPoverUB.tex}
    \caption{Numerical evaluation of the average ratio between the capacity gap $g_\text{TA}$, defined in~\eqref{eq:g_TA}, and the upper bound $\overline{\C}_{\text{TA}}$, derived from~\eqref{eq:P-SP}. The average ratio is plotted versus the SNR and for $\N=2,\dots,10$.}
    \label{fig:percMCgapTA}
\end{figure}%
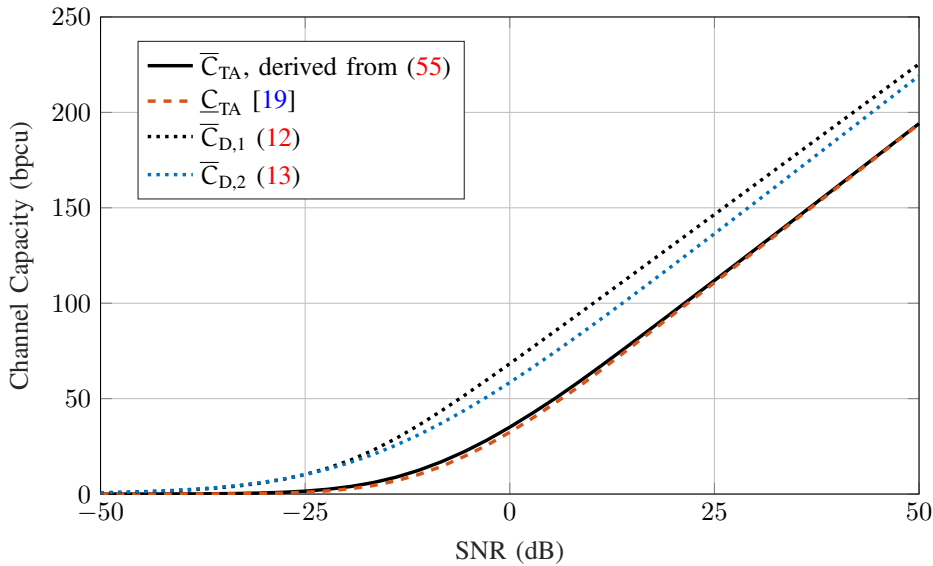
\begin{figure}[t]
    \centering
    %% ------------------------------------------------------------------------------------
%% ------------------------------------------------------------------------------------
%% ------------------------------------------------------------------------------------

% 6.8, 5.9, 5.2, 5, 3.7, 3.5, 2.4, 1.6, 1.1, 0.5 

\definecolor{mycolor1}{rgb}{0.00000,0.44700,0.74100}%
\definecolor{mycolor2}{rgb}{0.85000,0.32500,0.09800}%
\definecolor{mycolor3}{rgb}{0.92900,0.69400,0.12500}%
\definecolor{mycolor4}{rgb}{0.49400,0.18400,0.55600}%
\definecolor{mycolor5}{rgb}{0.46600,0.67400,0.18800}%
\definecolor{mycolor6}{rgb}{0.30100,0.74500,0.93300}%
\definecolor{mycolor7}{rgb}{0.63500,0.07800,0.18400}%
\definecolor{mycolor8}{rgb}{0.30000,0.85000,0.70862}%
\begin{tikzpicture}

\begin{axis}[%
width=0.75*0.8\linewidth,
height=0.7*0.5\linewidth,
at={(0\linewidth,0\linewidth)},
scale only axis,
xmin=-50,
xmax=50,
xtick={-50, -25,   0,  25,  50},
xlabel style={font=\color{white!15!black}},
xlabel={SNR (dB)},
ymin=0,
ymax=250,
ylabel style={font=\color{white!15!black}},
ylabel={Channel Capacity (bpcu)},
axis background/.style={fill=white},
xmajorgrids,
ymajorgrids,
legend style={at={(rel axis cs:0.445,0.95)},legend cell align=left, align=left, draw=white!15!black}
]
\addplot [color=black, line width=1.3pt]
  table[row sep=crcr]{%
-50	0.00720918469083642\\
-47.2222222222222	0.0136362637592526\\
-44.4444444444444	0.0257422729088432\\
-41.6666666666667	0.0484169217082879\\
-38.8888888888889	0.0904477847881751\\
-36.1111111111111	0.16691493092226\\
-33.3333333333333	0.30161325541355\\
-30.5555555555556	0.527742874095836\\
-27.7777777777778	0.906799623985165\\
-25	1.52810317394482\\
-22.2222222222222	2.50561316249808\\
-19.4444444444444	3.95583259938181\\
-16.6666666666667	6.01524465743174\\
-13.8888888888889	8.82759936895917\\
-11.1111111111111	12.4800229737349\\
-8.33333333333334	16.9508779296685\\
-5.55555555555556	22.2258739125224\\
-2.77777777777778	28.2801416353736\\
0	35.0597938515567\\
2.77777777777778	42.4062623830853\\
5.55555555555556	50.1663756504698\\
8.33333333333334	58.4390521232505\\
11.1111111111111	67.0551706446075\\
13.8888888888889	75.8618169966834\\
16.6666666666667	84.7990970026125\\
19.4444444444444	93.7932120564306\\
22.2222222222222	102.8348344966\\
25	111.894156093145\\
27.7777777777778	120.979204707569\\
30.5555555555556	130.079758142755\\
33.3333333333333	139.19687086503\\
36.1111111111111	148.330125634106\\
38.8888888888889	157.473622186296\\
41.6666666666667	166.625354420502\\
44.4444444444444	175.772477833884\\
47.2222222222222	184.943866156431\\
50	194.130123379862\\
};
\addlegendentry{$\overline{\C}_{\text{TA}}$, derived from~\eqref{eq:P-SP}}

\addplot [color=mycolor2, dashed, line width=1.3pt]
  table[row sep=crcr]{%
-50	0.00334593429064817\\
-47.2222222222222	0.00634081233037136\\
-44.4444444444444	0.0120126275578761\\
-41.6666666666667	0.0227445313309424\\
-38.8888888888889	0.0430166313123099\\
-36.1111111111111	0.0811884872440573\\
-33.3333333333333	0.152640997735731\\
-30.5555555555556	0.284941251788052\\
-27.7777777777778	0.5251615192283\\
-25	0.946915323093964\\
-22.2222222222222	1.64863923993017\\
-19.4444444444444	2.78553405282143\\
-16.6666666666667	4.48720689199375\\
-13.8888888888889	6.93283763886637\\
-11.1111111111111	10.2501672661552\\
-8.33333333333334	14.4760899630128\\
-5.55555555555556	19.7280163049096\\
-2.77777777777778	25.6051192648917\\
0	32.3782273831813\\
2.77777777777778	39.7885374032033\\
5.55555555555556	47.8037463980235\\
8.33333333333334	55.9531539084669\\
11.1111111111111	64.8321693671118\\
13.8888888888889	73.9833491857128\\
16.6666666666667	83.1704636018833\\
19.4444444444444	92.3766512824806\\
22.2222222222222	101.592933124645\\
25	110.814548858939\\
27.7777777777778	120.038980792067\\
30.5555555555556	129.26489898581\\
33.3333333333333	138.491601381092\\
36.1111111111111	147.718717497988\\
38.8888888888889	156.946051868367\\
41.6666666666667	166.173501371695\\
44.4444444444444	175.401011608817\\
47.2222222222222	184.62855388333\\
50	193.85611305765\\
};
\addlegendentry{$\underline{\C}_\text{TA}$~\cite{ourISIT2020}}

\addplot [color=black, dotted, line width=1.3pt]
  table[row sep=crcr]{%
-50	0.634142266022401\\
-47.2222222222222	0.871241113026265\\
-44.4444444444444	1.19601841818027\\
-41.6666666666667	1.64003638791301\\
-38.8888888888889	2.24545711460836\\
-36.1111111111111	3.0679278039476\\
-33.3333333333333	4.17964013262735\\
-30.5555555555556	5.67193996471366\\
-27.7777777777778	7.65621303664223\\
-25	10.2608081541402\\
-22.2222222222222	13.6207074735104\\
-19.4444444444444	17.8564524889432\\
-16.6666666666667	23.04136837917\\
-13.8888888888889	29.1633588013823\\
-11.1111111111111	36.1033321700343\\
-8.33333333333334	43.6656728320213\\
-5.55555555555556	51.6510614251207\\
-2.77777777777778	59.9071215806585\\
0	68.3340457617926\\
2.77777777777778	76.8702333127845\\
5.55555555555556	85.4778696278453\\
8.33333333333334	94.1333037195177\\
11.1111111111111	102.821352590486\\
13.8888888888889	111.532016786272\\
16.6666666666667	120.258564458001\\
19.4444444444444	128.99638407223\\
22.2222222222222	137.742276314985\\
25	146.494004265243\\
27.7777777777778	155.250000579862\\
30.5555555555556	164.009173500234\\
33.3333333333333	172.770777303003\\
36.1111111111111	181.534326393351\\
38.8888888888889	190.299540271426\\
41.6666666666667	199.066311563948\\
44.4444444444444	207.834692530094\\
47.2222222222222	216.604897638351\\
50	225.37732133425\\
};
\addlegendentry{$\overline{\C}_{\text{D,1}}$~\eqref{eq:DUBball}}

\addplot [color=mycolor1, dotted, line width=1.3pt]
  table[row sep=crcr]{%
-50	0.709098274769803\\
-47.2222222222222	0.970526243877599\\
-44.4444444444444	1.32550865779743\\
-41.6666666666667	1.80519655790847\\
-38.8888888888889	2.44929432563258\\
-36.1111111111111	3.30708613705905\\
-33.3333333333333	4.43766700733593\\
-30.5555555555556	5.90879491454535\\
-27.7777777777778	7.79382334849446\\
-25	10.1665067732411\\
-22.2222222222222	13.0941195752293\\
-19.4444444444444	16.6300907585613\\
-16.6666666666667	20.8078263925649\\
-13.8888888888889	25.6372156354816\\
-11.1111111111111	31.1045058977528\\
-8.33333333333334	37.17520739527\\
-5.55555555555556	43.798961536069\\
-2.77777777777778	50.9151312538393\\
0	58.4581395060986\\
2.77777777777778	66.3620090595878\\
5.55555555555556	74.5638994993273\\
8.33333333333334	83.0066131927854\\
11.1111111111111	91.6401093458617\\
13.8888888888889	100.422117138658\\
16.6666666666667	109.318008422271\\
19.4444444444444	118.300148556232\\
22.2222222222222	127.346956475601\\
25	136.441869828021\\
27.7777777777778	145.572350787984\\
30.5555555555556	154.729007820059\\
33.3333333333333	163.904862316315\\
36.1111111111111	173.094759769834\\
38.8888888888889	182.294909996424\\
41.6666666666667	191.502535232125\\
44.4444444444444	200.715604653783\\
47.2222222222222	209.932636177099\\
50	219.152549597038\\
};
\addlegendentry{$\overline{\C}_{\text{D,2}}$~\eqref{eq:DUBbox}}

\end{axis}

\end{tikzpicture}%
    \caption{Capacity bounds in bit per channel use (bpcu) versus SNR, for $\N=10$ and for a random realization of the complex matrix $\widetilde{\Hv}$.}
    \label{fig:TA_SPvsDytso}
\end{figure}%
Let us evaluate the SP upper bound of Section~\ref{S:SPbound} for the TA input constraint region ${\sX} = \ball[2 \N ]{\A}$, with constraint amplitude $\A \in \mathbb{R}_+$. As shown in~\eqref{eq:finalL}, to evaluate the bound $\overline{\C}_\text{SP}$ we need to be able to compute the intrinsic volumes $\iV[j]{\Hv \sX}$, for $j=0,\dots,2 \N $. As mentioned in Section~\ref{sss:P-SP}, when $\sX$ is a ball, by considering the singular value decomposition of $\Hv = \Uv \Dv \Vv^T$, it holds
\begin{align}
    \iV[j]{\Hv \sX} = \iV[j]{\Dv \sX}, \quad \forall j.
\end{align}
In~\cite{iV_Ellipsoid}, it is shown that given an ellipsoid
\begin{align}
\sE = \lrc{\xv = \lrs{x_1, \dots, x_{2 \N }}^T \in \mathbb{R}^{2 \N } : \xv^T \mathsf{\Sigma}^{-1} \xv \leq 1},
\end{align}
by defining $j$ independent and identically distributed random vectors $\Qv_1,\dots,\Qv_j \sim {\cal N} \lr{\textbf{\textsf{0}}_{2 \N } , \mathsf{\Sigma} }$ and the random matrix $\QMv = \lrs{\Qv_1,\dots,\Qv_j}$, it is possible to compute the $j$th intrinsic volume of $\sE$ as
\begin{align} \label{eq:iVellipsoid}
    \iV[j]{\sE} = \frac{(2 \pi)^{j/2}}{j!}\expect{\sqrt{\det \lr{\QMv^T \cdot \QMv } } }.
\end{align} 
Let us set $\mathsf{\Sigma} = \Dv^2$. Then, the intrinsic volumes for the TA configuration are given by
\begin{align} \label{eq:intrV}
    \iV[j]{\Hv \sX} \stackrel{\eqref{eq:homogeneous}}{=} \iV[j]{\sE} \A^j, \qquad j=0,\ldots, 2 \N .
\end{align}
By plugging the intrinsic volumes~\eqref{eq:intrV} into~\eqref{eq:finalL}, we obtain the SP upper bound on the capacity for the TA constraint. Finally, to improve the performance of the SP bound at low SNR, we can apply the P-SP upper bound defined in Section~\ref{sss:P-SP}.

\subsubsection*{Capacity Gap and Performance}
To evaluate the gap we consider the piecewise-EPI (P-EPI) lower bound proposed in~\cite{ourISIT2020} and we denote it with $\underline{\C}_\text{TA}$. Let us define the gap as 
\begin{align} \label{eq:g_TA}
    g_\text{TA} \triangleq \overline{\C}_\text{TA} - \underline{\C}_\text{TA},
\end{align}
where $\overline{\C}_\text{TA}$ is given by the P-SP upper bound in~\eqref{eq:P-SP}, applied to the TA constraint. We evaluate $g_\text{TA}$ numerically by Monte Carlo simulation for $\N=2,\dots,10$, over random channel realizations. The entries of $\widetilde{\Hv}$ in~\eqref{eq:model} are drawn independently as $\widetilde{H}_{i,j} \sim {\cal CN}(0,2), \ \forall i,j$. 

The results are presented in Figs.~\Cref{fig:MCgapTA,fig:MCgap/N_TA,fig:percMCgapTA}. In Fig.~\ref{fig:MCgapTA}a, we show a scatter plot of the gap realizations and with the solid lines the average gap, both versus SNR and for each $\N=2,\dots,10$. In Fig.~\ref{fig:MCgapTA}b, it is shown the standard deviation of $g_\text{TA}$ for each $\N$. As expected, when the SNR goes to zero the gap is small, because~\eqref{eq:P-SP} is optimized by $\U=0$ and $\overline{\C}_\text{TA}$ turns into the Gaussian maximizing entropy bound, which is tight at low SNR. Moreover, the gap is vanishing at high SNR as proven in Section~\ref{S:SPbound}. Even in the worst case, $g_\text{TA}$ is approximately $3 \text{ bit per channel use}$~(bpcu). In both Fig.~\ref{fig:MCgap/N_TA} and Fig.~\ref{fig:percMCgapTA} we show that as $\N$ increases the performance of the upper bound improves. In Fig.~\ref{fig:MCgap/N_TA}, it is shown the average gap per complex dimension $\N$ in solid lines, while in Fig.~\ref{fig:percMCgapTA} we show the ratio between the bounding gap and the upper bound. In Fig.~\ref{fig:MCgap/N_TA}, we also compare the gap of the proposed bound with the one resulting from the duality upper bounds of~\cite{Dytso} defined in~\eqref{eq:DUBball} and~\eqref{eq:DUBbox}. Given $\overline{\C}_{\text{D,TA}} \triangleq \min \lr{\overline{\C}_{\text{D,1}},\overline{\C}_{\text{D,2}}} $, the capacity gap for the duality upper bounds is given by
\begin{align} \label{eq:gD_TA}
    g_{\text{D,TA}} \triangleq \overline{\C}_{\text{D,TA}} - \underline{\C}_\text{TA}. 
\end{align}
The dashed lines in Fig.~\ref{fig:MCgap/N_TA} are the average gaps for complex dimension $\expect{g_{\text{D,TA}}}/\N$, which are larger than $\expect{g_{\text{TA}}}/\N$ for any SNR level. Finally, in Fig.~\ref{fig:TA_SPvsDytso} we show the capacity bounds for a random channel realization given $\N=10$. The figure shows both how close our P-SP bound is to the lower bound and also the substantial improvement compared to $\overline{\C}_{\text{D,1}}$ and $\overline{\C}_{\text{D,2}}$.

\subsection{Per-Antenna Constraint}
\label{S:PAbound}

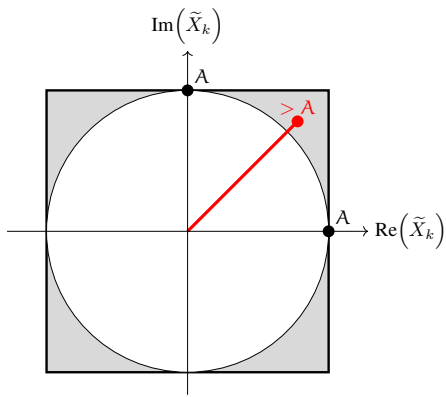
\begin{figure}[t]
	\centering
	\scalebox{0.75}{%
	\begin{tikzpicture}
	\filldraw[fill=black!15,draw=black,very thick] (-2.5,-2.5) rectangle (2.5,2.5);
	\filldraw[fill=white,draw=black] (0,0) circle (2.5);
	\draw[red,ultra thick,-] (0,0)--(1.95,1.95);
	\draw[->] (-3.2,0)--(3.2,0) node[right] {$\Re{ \widetilde{X}_k }$};
	\draw[->] (0,-2.9)--(0,3.2) node[above] {$\Im{ \widetilde{X}_k }$};
	\fill[fill=black] (2.5,0) circle (3pt) node[anchor=south west] {$\A$};
	\fill[fill=black] (0,2.5) circle (3pt) node[anchor=south west] {$\A$};
	\fill[fill=red] (1.95,1.95) circle (3pt)
	node[above,text=red] {$>\A$};
	\node[label=left:{$\quad$}] at (-3.15,0) {};;
	\end{tikzpicture}
	}
	\caption{Resulting constraint region for the complex input (circle) and for the real and imaginary part of the signal, independently (square).}
	\label{fig:Comp_Const_vs_ReIm_Const}
\end{figure}
\begin{figure}[t]
    \centering
    \scalebox{1}{\begin{tikzpicture}

\fill[black] (0,0,0) circle(1.75pt);

\draw[-Stealth,thick] (0,0,0) -- (3.2,-0.3,0);
\node[below] at (1.6,-0.15,0) {$\rv_1$};
\draw[-Stealth,thick] (0,0,0) -- (0,-0.1,-3);
\node[right] at (0.075,-0.1,-1.5) {$\rv_2$};
\draw[-Stealth,thick] (0,0,0) -- (0.5,2,0);
\node[left] at (0.25,1,0) {$\rv_3$};

\node[left] at (0,0,0) {$\p$};

\draw[] (3.2,-0.3,0) -- ++(0.5,2,0) -- ++(-3.2,0.3,0) -- ++(0,-0.1,-3) -- ++(3.2,-0.3,0) -- ++(0,0.1,3);

\draw[] (3.2,-0.3,0) -- ++(0,-0.1,-3) -- ++(-3.2,0.3,0) -- ++(0.5,2,0);
\draw[] (3.2,-0.3,0)++(0,-0.1,-3) -- ++(0.5,2,0);

\end{tikzpicture}%
}
    \caption{Parallelepiped spanned by $\rv_1$, $\rv_2$, and $\rv_3$ in $\p$.}
    \label{fig:3Dparall}
\end{figure}
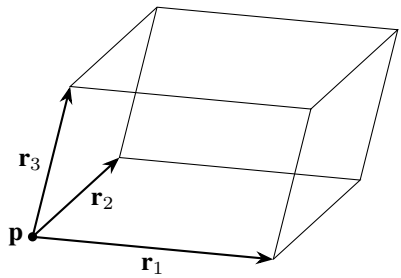
\begin{figure}[t]
    \centering
    \input{Figures/PA_CapacityGapANDsigma}
    \caption{\textbf{a)} Numerical evaluation of the capacity gap $g_\text{PA}$, defined in~\eqref{eq:g_PA}, in bit per channel use (bpcu) versus SNR, for $\N=2,\dots,10$. For each $\N$, the filled circles are the gaps resulting from each random channel realization, while the solid lines show the averaged behavior. \textbf{b)} Standard deviation of $g_\text{PA}$ in bpcu versus SNR, for $\N=2,\dots,10$.}
    \label{fig:MCgapPA}
\end{figure}
\begin{figure}[t]
    \centering
    \input{Figures/PA_CapacityGap_overN}
    \caption{Numerical evaluation of the average capacity gap per complex dimension in bit per channel use (bpcu) versus SNR, for $\N=2,\dots,10$. The solid lines are $\expect{g_\text{PA}}/\N$, with $g_\text{PA}$ defined in~\eqref{eq:g_PA}. The dashed lines are $\expect{g_\text{D,PA}}/\N$, with $g_\text{D,PA}$ defined in~\eqref{eq:gD_PA}.}
    \label{fig:MCgapPA_N}
\end{figure}
\begin{figure}[t]
    \centering
    \input{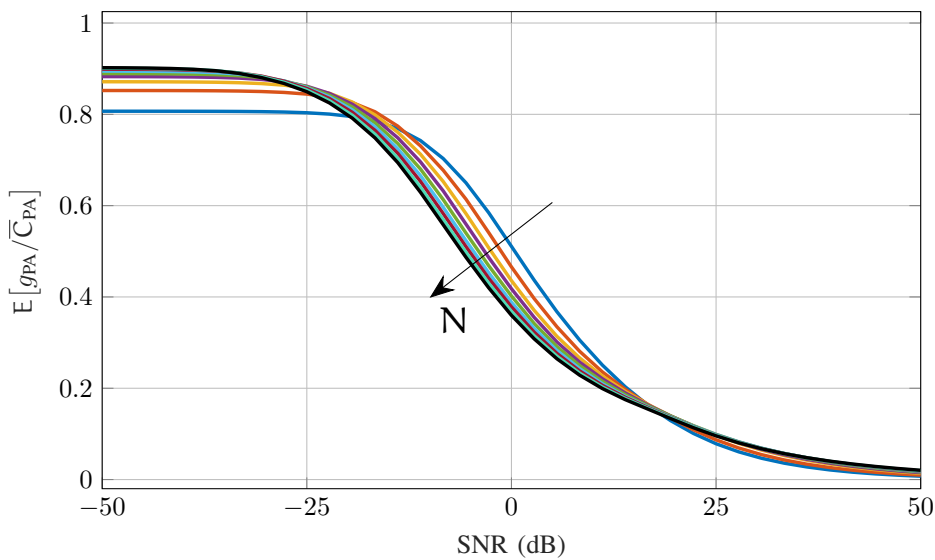}
    \caption{Numerical evaluation of the average ratio between the capacity gap $g_\text{PA}$, defined in~\eqref{eq:g_PA}, and the upper bound $\overline{\C}_{\text{PA}}$, defined in~\eqref{eq:C_PA_UB}. The average ratio is plotted versus the SNR and for $\N=2,\dots,10$.}
    \label{fig:percMCgapPA_UB}
\end{figure}
\begin{figure}[t]
    \centering
    %% ------------------------------------------------------------------------------------
%% ------------------------------------------------------------------------------------
%% ------------------------------------------------------------------------------------

% 6.8, 5.9, 5.2, 5, 3.7, 3.5, 2.4, 1.6, 1.1, 0.5 

\definecolor{mycolor1}{rgb}{0.00000,0.44700,0.74100}%
\definecolor{mycolor2}{rgb}{0.85000,0.32500,0.09800}%
\definecolor{mycolor3}{rgb}{0.92900,0.69400,0.12500}%
\definecolor{mycolor4}{rgb}{0.49400,0.18400,0.55600}%
\definecolor{mycolor5}{rgb}{0.46600,0.67400,0.18800}%
\definecolor{mycolor6}{rgb}{0.30100,0.74500,0.93300}%
\definecolor{mycolor7}{rgb}{0.63500,0.07800,0.18400}%
\definecolor{mycolor8}{rgb}{0.30000,0.85000,0.70862}%
\begin{tikzpicture}

\begin{axis}[%
width=0.75*0.8\linewidth,
height=0.7*0.5\linewidth,
at={(0\linewidth,0\linewidth)},
scale only axis,
xmin=-50,
xmax=50,
xtick={-50, -25,   0,  25,  50},
xlabel style={font=\color{white!15!black}},
xlabel={SNR (dB)},
ymin=0,
ymax=250,
ylabel style={font=\color{white!15!black}},
ylabel={Channel Capacity (bpcu)},
axis background/.style={fill=white},
xmajorgrids,
ymajorgrids,
legend style={at={(rel axis cs:0.445,0.95)},legend cell align=left, align=left, draw=white!15!black}
]
\addplot [color=black, line width=1.3pt]
  table[row sep=crcr]{%
-50	0.352935945716304\\
-47.2222222222222	0.488844517287702\\
-44.4444444444444	0.67376443789249\\
-41.6666666666667	0.925752436109826\\
-38.8888888888889	1.27042050470091\\
-36.1111111111111	1.73776131548948\\
-33.3333333333333	2.36018529916544\\
-30.5555555555556	3.18532499615059\\
-27.7777777777778	4.2737569257939\\
-25	5.68095072068446\\
-22.2222222222222	7.48911866056379\\
-19.4444444444444	9.76437872873778\\
-16.6666666666667	12.5477233980315\\
-13.8888888888889	15.876788592348\\
-11.1111111111111	19.7686617531018\\
-8.33333333333334	24.2965402973771\\
-5.55555555555556	29.4054532170124\\
-2.77777777777778	35.0108037457395\\
0	41.1224192896983\\
2.77777777777778	47.6516557682151\\
5.55555555555556	54.4970904131573\\
8.33333333333334	61.6056979811526\\
11.1111111111111	68.9475788328815\\
13.8888888888889	76.4659398122755\\
16.6666666666667	84.1352754633277\\
19.4444444444444	91.9459315813916\\
22.2222222222222	99.8943422067221\\
25	107.968944820409\\
27.7777777777778	116.158015411214\\
30.5555555555556	124.447743153295\\
33.3333333333333	132.841468991822\\
36.1111111111111	141.320419839036\\
38.8888888888889	149.875894137225\\
41.6666666666667	158.491185650769\\
44.4444444444444	167.167606243644\\
47.2222222222222	175.888036513854\\
50	184.651445338265\\
};
\addlegendentry{$\overline{\C}_{\text{PA,1}}$, derived from~\eqref{eq:G-SP}}

\addplot [color=mycolor1, line width=1.3pt]
  table[row sep=crcr]{%
-50	0.00664607950273677\\
-47.2222222222222	0.01257332611479\\
-44.4444444444444	0.0237434140041586\\
-41.6666666666667	0.0446844629818628\\
-38.8888888888889	0.0835677011853022\\
-36.1111111111111	0.15452178786142\\
-33.3333333333333	0.280146119518726\\
-30.5555555555556	0.494026610235075\\
-27.7777777777778	0.857798825772562\\
-25	1.4471152260255\\
-22.2222222222222	2.39818828474701\\
-19.4444444444444	3.85887265729124\\
-16.6666666666667	5.90929470081325\\
-13.8888888888889	8.74450492323961\\
-11.1111111111111	12.4231473253576\\
-8.33333333333334	16.8351046187033\\
-5.55555555555556	21.9881385184556\\
-2.77777777777778	27.8938309597493\\
0	34.5186822225782\\
2.77777777777778	41.7635859039628\\
5.55555555555556	49.4198821162679\\
8.33333333333334	57.590643786869\\
11.1111111111111	66.1449347985593\\
13.8888888888889	74.9309507703149\\
16.6666666666667	83.842783723516\\
19.4444444444444	92.8024407011811\\
22.2222222222222	101.831121525647\\
25	110.906512158789\\
27.7777777777778	119.816758477283\\
30.5555555555556	128.439036427921\\
33.3333333333333	137.101276407315\\
36.1111111111111	145.782635327472\\
38.8888888888889	154.499992988563\\
41.6666666666667	163.217912409454\\
44.4444444444444	171.934910891611\\
47.2222222222222	180.690125177911\\
50	189.454054591436\\
};
\addlegendentry{$\overline{\C}_{\text{PA,2}}$, derived from~\eqref{eq:P-SP}}

\addplot [color=mycolor2, dashed, line width=1.3pt]
  table[row sep=crcr]{%
-50	0.000759367709204071\\
-47.2222222222222	0.00143950388467158\\
-44.4444444444444	0.00272871561880692\\
-41.6666666666667	0.00517219336923704\\
-38.8888888888889	0.00980248890453286\\
-36.1111111111111	0.0185735127657655\\
-33.3333333333333	0.0351767009006305\\
-30.5555555555556	0.0665648129084754\\
-27.7777777777778	0.125757773421712\\
-25	0.23687295040075\\
-22.2222222222222	0.443681472973038\\
-19.4444444444444	0.822670496847526\\
-16.6666666666667	1.49854588228384\\
-13.8888888888889	2.65086386805927\\
-11.1111111111111	4.48693948199238\\
-8.33333333333334	7.26254431001392\\
-5.55555555555556	11.0799032046903\\
-2.77777777777778	15.8056712012435\\
0	21.6776550245989\\
2.77777777777778	28.4810002777531\\
5.55555555555556	36.0829178117073\\
8.33333333333334	44.00103566532\\
11.1111111111111	52.6382120654153\\
13.8888888888889	61.6872676699493\\
16.6666666666667	70.8197775789835\\
19.4444444444444	79.9969536758208\\
22.2222222222222	89.1978735956883\\
25	98.4113696153226\\
27.7777777777778	107.631513852956\\
30.5555555555556	116.85516902129\\
33.3333333333333	126.080677318456\\
36.1111111111111	135.307163450719\\
38.8888888888889	144.534165477052\\
41.6666666666667	153.761439661386\\
44.4444444444444	162.988857415672\\
47.2222222222222	172.216350904927\\
50	181.443884344875\\
};
\addlegendentry{$\underline{\C}_\text{PA}$~\cite{ourISIT2020}}

\addplot [color=black, dotted, line width=1.3pt]
  table[row sep=crcr]{%
-50	0.608952406034558\\
-47.2222222222222	0.836705011561851\\
-44.4444444444444	1.14874393152804\\
-41.6666666666667	1.5754672843498\\
-38.8888888888889	2.15753314338044\\
-36.1111111111111	2.94869935613281\\
-33.3333333333333	4.01888521445783\\
-30.5555555555556	5.45689260365767\\
-27.7777777777778	7.37161303899289\\
-25	9.88961927217064\\
-22.2222222222222	13.1459675228009\\
-19.4444444444444	17.2646378009468\\
-16.6666666666667	22.3271098193971\\
-13.8888888888889	28.3340704889977\\
-11.1111111111111	35.1797173165869\\
-8.33333333333334	42.6744418944862\\
-5.55555555555556	50.6155689443101\\
-2.77777777777778	58.8436635265622\\
0	67.2528736710366\\
2.77777777777778	75.7776203008014\\
5.55555555555556	84.377692207887\\
8.33333333333334	93.0280156447216\\
11.1111111111111	101.71254888228\\
13.8888888888889	110.420759569954\\
16.6666666666667	119.145575182812\\
19.4444444444444	127.882160272834\\
22.2222222222222	136.627164707308\\
25	145.378247763394\\
27.7777777777778	154.133769196928\\
30.5555555555556	162.892584936041\\
33.3333333333333	171.653910690237\\
36.1111111111111	180.417231321533\\
38.8888888888889	189.182242392636\\
41.6666666666667	197.948815568985\\
44.4444444444444	206.716982941771\\
47.2222222222222	215.486937631809\\
50	224.259049630418\\
};
\addlegendentry{$\overline{\C}_{\text{D,1}}$~\eqref{eq:DUBball}}

\addplot [color=mycolor1, dotted, line width=1.3pt]
  table[row sep=crcr]{%
-50	0.695189547932158\\
-47.2222222222222	0.951562698541847\\
-44.4444444444444	1.29974088015842\\
-41.6666666666667	1.77033822427162\\
-38.8888888888889	2.40240667438207\\
-36.1111111111111	3.24446915511063\\
-33.3333333333333	4.3547773690588\\
-30.5555555555556	5.80021504300895\\
-27.7777777777778	7.65330624383442\\
-25	9.98711922442233\\
-22.2222222222222	12.8685072260839\\
-19.4444444444444	16.3508919146538\\
-16.6666666666667	20.468250120815\\
-13.8888888888889	25.2317486300548\\
-11.1111111111111	30.629626561506\\
-8.33333333333334	36.6299204949397\\
-5.55555555555556	43.1849868804769\\
-2.77777777777778	50.2366879546504\\
0	57.7213968881054\\
2.77777777777778	65.5743552641505\\
5.55555555555556	73.7331896209924\\
8.33333333333334	82.1405233079908\\
11.1111111111111	90.7456691440786\\
13.8888888888889	99.5054398339564\\
16.6666666666667	108.384194107111\\
19.4444444444444	117.353313781319\\
22.2222222222222	126.390338790686\\
25	135.477964851208\\
27.7777777777778	144.603052975053\\
30.5555555555556	153.755738748867\\
33.3333333333333	162.928679547293\\
36.1111111111111	172.1164450322\\
38.8888888888889	181.31503840404\\
41.6666666666667	190.521528425537\\
44.4444444444444	199.733770969951\\
47.2222222222222	208.950200677416\\
50	218.169676335594\\
};
\addlegendentry{$\overline{\C}_{\text{D,2}}$~\eqref{eq:DUBbox}}

\end{axis}

\end{tikzpicture}%
    \caption{Capacity bounds in bit per channel use (bpcu) versus SNR, for $\N=10$ and for a random realization of the complex matrix $\widetilde{\Hv}$.}
    \label{fig:PA_SPvsDytso}
\end{figure}
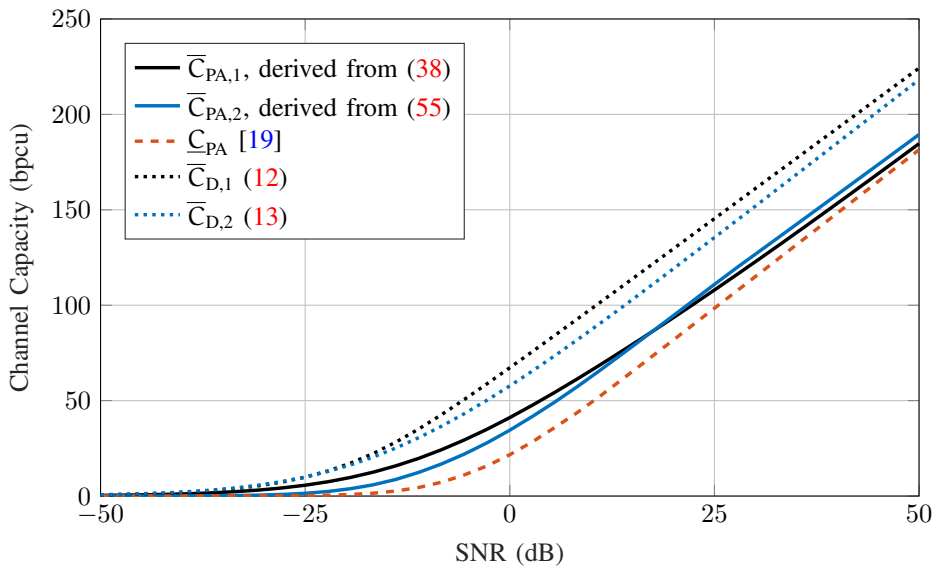

Let us now consider the PA constraint. The complex input is such that $\widetilde{
\Xv} \in \widetilde{\sX} = \Bocs[\N]{2\bm{\A}}$, with $\bm{\A}=\mleft( \A_1,\dots,\A_{\N} \mright) \in {\mathbb{R}}^{\N}_+$ being the vector of amplitude constraints. Since we can always consider an equivalent system by absorbing unequal $\A_k$'s in the channel matrix, we can assume $\A_k=\A, \, \forall k$ without loss of generality. Notice that constraining each entry of the vectorized input vector does not induce the same constraint as $\widetilde{\sX}$. Applying the constraint on the entries of $\Xv$, and therefore after the vectorization of~\eqref{eq:model} like in~\cite{Dytso}, increases the capacity. Indeed, Fig.~\ref{fig:Comp_Const_vs_ReIm_Const} shows that the constraint on the real and imaginary parts of $\widetilde{X}_k$ is weaker if compared to $\big|\widetilde{X}_k\big| \leq \A$, therefore it would induce a larger capacity. We think that applying the constraint on $\widetilde{\Xv}$, instead of $\Xv$, correctly interprets the technological limitations imposed by power amplifiers. Then, the equivalent $\N$ real $2$-dimensional constraints induced by $\big|\widetilde{X}_k\big| \leq \A$ are
\begin{align}
    \lrs{\Re{\widetilde{X}_k},\Im{\widetilde{X}_k}} \in \sX_k = \ball[2]{\A}, \ \forall k=1,\dots,\N.
\end{align}
We remark that, given the PA constraint, the set $\sX$ is then defined as the Cartesian product of $\N$ circles, $\sX \triangleq \sX_1 \times \dots \times \sX_\N$. Although it is fairly easy to evaluate the intrinsic volumes $\iV[j]{\sX}$, the channel matrix $\Hv$ can distort $\sX$ in such a way that $\Hv \sX$ is not anymore a Cartesian product. Because of this distortion, even when the intrinsic volumes of $\sX$ are known, it is not trivial to evaluate those of $\Hv \sX$. Therefore, the only viable solution is to apply the G-SP bound by deriving upper bounds on $\iV[j]{\Hv \sX}$.

Let us substitute each $\sX_k$ with a larger region $\sR_k = \Bocs[2]{2 \A} \supset \sX_k, \ \forall k=1,\dots,\N$. Then we have
\begin{align}
\sR = \sR_1 \times \dots \times \sR_\N = \Bocs[2 \N ]{2 \A} \supset \sX.
\end{align}
We now show that it is possible to evaluate an upper bound on the intrinsic volumes $\iV[j]{\Hv \sR}$ and therefore to derive a G-SP bound for the PA constraint. Let $\sP$ be a $2 \N $-dimensional polytope and let $\mathbb{F}_j(\sP)$ denote the set of all $j$-dimensional faces of $\sP$. From~\cite{mcmullen1975non}, the intrinsic volumes of a polytope are defined as
\begin{align} \label{eq:polyiV}
     \iV[j]{\sP} = \sum_{\sF\in \mathbb{F}_j(\sP) } \gamma(\sF,\sP) \Vol[j]{\sF},
\end{align}
where $\gamma(\sF,\sP)$ is the normalized external angle\footnote{For a more rigorous definition of external angle, see~\cite{grunbaum1968grassmann}.} of $\sP$ at its face $\sF$. Note that, since $\gamma(\sF,\sP) \leq 1$, we can upper-bound and simplify~\eqref{eq:polyiV} with
    \begin{align} \label{eq:polyUBiV}
     \iV[j]{\sP} \leq \sum_{\sF\in \mathbb{F}_j(\sP) } \Vol[j]{\sF}.
\end{align}
Notice that $\sR$ is a parallelepiped, therefore by using~\eqref{eq:polyUBiV} and results from exterior algebra, it is possible to evaluate an upper bound on $\iV[j]{\Hv \sR}$. The authors of~\cite{wedgematrixgeo} show how a $j$-dimensional parallelepiped, with $j\leq 2\N$, can be identified by a set of $j$ vectors $\rv_1, \dots, \rv_j \in \mathbb{R}^{2\N}$ and by a base point $\p \in \mathbb{R}^{2\N}$. Let $\p$ be one of the vertices of the parallelepiped and let $\rv_1, \dots, \rv_j$ have the same magnitude and direction of the $j$ edges originating from $\p$. Then, the parallelepiped is composed of all points in $\mathbb{R}^{2\N}$ resulting from
\begin{align}
    \p + t_1 \rv_1 + \dots +t_j \rv_j, \quad 0 \leq t_1, \dots, t_j \leq 1.
\end{align}
For instance, in Fig.~\ref{fig:3Dparall} it is shown the $3$-dimensional parallelepiped spanned by linear combinations of the vectors $\rv_1$, $\rv_2$, and $\rv_3$ given the base point $\p$.
Since volume is invariant with respect to translations, we can drop the base point $\p$ and represent the geometric region $\sR$ via the corresponding matrix $\Rv$. Let us define it as
\begin{align}
    \Rv \triangleq 
    \begin{bmatrix}
    \rv_1 & \rv_2 & \dots & \rv_{2 \N }
    \end{bmatrix}
    = 2 \A \cdot \Iv_{2 \N }. 
\end{align}
To evaluate the $j$th intrinsic volumes $\iV[j]{\sR}$, with $j \leq 2 \N $, we need to compute the $j$-dimensional volumes of all the faces spanned by all the possible combinations of $j$ column vectors in $\Rv$, accounting for all possible repetitions. Let us denote by $\sR_{j,i}$ the $j$-dimensional face spanned by the $i$th combination of $j$ column vectors in $\Rv$, out of $\binom{2\N}{j}$. For instance, let us consider $\sR_{j,1}$ to be the face spanned by the $\rv_k$'s with $k = 1, \dots , j$. Then, the corresponding $2\N \times j$~matrix is
$
    \Rv_{j,1} = 
    \begin{bmatrix}
    \rv_1 & \rv_2 & \dots & \rv_j
    \end{bmatrix}
$. As shown in~\cite{wedgematrixgeo}, the $j$-dimensional volume of the face $\sR_{j,1}$ is given by
\begin{align}
    \Vol[j]{\sR_{j,1}} = \sqrt{ \abs{ \det \lr{\Rv_{j,1}^T \cdot \Rv_{j,1} } } }.
\end{align}
The same reasoning can be extended to any $i$th combination and it can be applied analogously after the distortion introduced by the channel matrix $\Hv$. Let us define the region $\sS = \Hv \sR \supset \Hv \sX$ and the corresponding $2\N \times 2 \N $ matrix $\Sv = \begin{bmatrix}
    \sv_1 & \sv_2 & \dots & \sv_{2 \N }
    \end{bmatrix} = \Hv \cdot \Rv$.
Let $\sSbox$ be the region characterized by the same intrinsic volumes of $\sS$, but with external angles always equal to $1$. Then by~\eqref{eq:polyUBiV}, the $j$th intrinsic volume of $\sS$ can be upper-bounded by
\begin{align} \label{eq:iVUBbox}
     \iV[j]{\sS} \leq \iV[j]{\sSbox} = 2^{2 \N -j} \sum_{i=1}^{\binom{2 \N }{j}} \sqrt{ \abs{ \det \lr{\Sv_{j,i}^T \cdot \Sv_{j,i} } } },
\end{align}
where the term $2^{2 \N -j}$ accounts for how many times each $j$-dimensional face is repeated in any parallelepiped. Since $\iV[j]{\Hv \sX} \leq\iV[j]{\sS} \leq \iV[j]{\sSbox}$,~\eqref{eq:iVUBbox} provides us with all the elements needed to apply the G-SP of Section~\ref{S:GA}.

Another suitable upper bound on the intrinsic volumes of $\Hv \sX$ can be derived by considering a region $\sSball \triangleq \Hv \ball[2 \N]{ r_{\text{max}} \lr{\sX} } = \Hv  \ball[2 \N]{\A \sqrt{\N}} \supset \Hv \sX$. Since $\sSball$ is an ellipsoid, its intrinsic volumes are easily derived as
\begin{align} \label{eq:iVUBball}
    \iV[j]{ \sSball } = \iV[j]{\sE}\lr{ \A \sqrt{\N}}^j,
\end{align}
where $\iV[j]{\sE}$ is defined in~\eqref{eq:iVellipsoid}, with $\mathsf{\Sigma} = \Dv^2$ and $\Hv = \Uv \Dv \Vv^T$. 
Since both $\sSball$ and $\sSbox$ can provide valid upper bounds, we choose the best option between~\eqref{eq:iVUBbox} and~\eqref{eq:iVUBball} for each $j$, \emph{i.e.},
\begin{align}\label{eq:iVUBmin}
    \iV[j]{\sS_j} = \min \lr{\iV[j]{ \sSbox }, \iV[j]{ \sSball }}.
\end{align} %
Then, by plugging~\eqref{eq:iVUBmin} into~\eqref{eq:geniVUB} and, in turn, into~\eqref{eq:G-SP} we obtain the G-SP upper bound for the PA constraint, namely $\overline{\C}_{\text{PA,1}}$. Notice that the true $2 \N $th intrinsic volume can be easily computed and it is given by $\iV[2 \N ]{\Hv \sX} = \Vol{\Hv \sX} = \det \lr{\Hv} \pi^\N \A^{2 \N }$. Finally, by considering $\ball[2\N]{r_{\text{max}} \lr{\sX}}  =  \ball[2\N]{ \A \sqrt{\N} } \supset \sX$, we can also evaluate the P-SP bound in~\eqref{eq:P-SP} for the PA constraint, which we will denote by $\overline{\C}_{\text{PA,2}}$.

\subsubsection*{Capacity Gap and Performance}
Let us define the capacity gap for the PA constraint as \begin{align} \label{eq:g_PA}
    g_{\text{PA}} \triangleq \overline{\C}_{\text{PA}} - \underline{\C}_{\text{PA}},
\end{align}
where $\underline{\C}_\text{PA}$ is the P-EPI lower bound for the PA constraint from~\cite{ourISIT2020} and $\overline{\C}_{\text{PA}}$ is given by
\begin{align} \label{eq:C_PA_UB}
    \overline{\C}_{\text{PA}} = \min \lr{\overline{\C}_{\text{PA,1}},\overline{\C}_{\text{PA,2}}}.
\end{align}
In Fig.~\ref{fig:MCgapPA}a, it is shown a scatter plot of the gap realizations and, with solid lines, the averaged behavior. Both are shown versus the SNR, for $\N = 2, \dots,10$, and evaluated over random channel matrix realizations. The entries of $\widetilde{\Hv}$ in~\eqref{eq:model} are drawn independently as $\widetilde{H}_{i,j} \sim {\cal CN}(0,2), \ \forall i,j$. In Fig.~\ref{fig:MCgapPA}b, we show the standard deviation of $g_{\text{PA}}$ for each $\N$. Fig.~\ref{fig:MCgapPA}a shows that, as for the TA constraint, the gap is small for SNR going to zero thanks to the P-SP approach and tends to decrease at high SNR. 
Notice that the capacity gap for the PA constraint, shown in Fig.~\ref{fig:MCgapPA}a, is larger than the one resulting from the TA constraint, shown in Fig.~\ref{fig:MCgapTA}a. This does not come as a surprise, it is simply due to the infeasibility in the direct evaluation of the intrinsic volumes of $\Hv \sX$ and the consequent necessity to upper-bound them via the G-SP and P-SP approaches. Nevertheless, in Fig.~\ref{fig:percMCgapPA_UB} we show that the average ratio between the gap $g_{\text{PA}}$ and the upper bound $\overline{\C}_{\text{PA}}$ is within $\approx 0.1$ after an SNR of $25$ dB. Furthermore, Fig.~\ref{fig:MCgapPA_N} and Fig.~\ref{fig:PA_SPvsDytso} clearly show the benefits that the presented upper bounds provide if compared to the duality upper bounds of~\cite{Dytso}. As for the TA constraint, the upper bounds $\overline{\C}_{\text{D,1}}$ and $\overline{\C}_{\text{D,2}}$ are derived from~\mbox{\cite[Theorem~10]{Dytso}} and defined in~\eqref{eq:DUBball} and~\eqref{eq:DUBbox} respectively. We remark that~\eqref{eq:DUBbox} requires the evaluation of the smallest box containing the region $\Hv \sX$, which is not always a trivial task. Therefore, as we did for $\overline{\C}_{\text{PA,2}}$, in the derivation of $\overline{\C}_{\text{D,2}}$ for the PA constraint we made the simplifying assumption of considering an input constraint region $\ball[2 \N]{ r_{\text{max}} \lr{\sX} } = \ball[2 \N]{ \A \sqrt{\N} } \supset \sX$. Then, we define $\overline{\C}_{\text{D,PA}} \triangleq \min \lr{\overline{\C}_{\text{D,1}},\overline{\C}_{\text{D,2}}}$ and the capacity gap
\begin{align} \label{eq:gD_PA}
g_{\text{D,PA}} \triangleq \overline{\C}_{\text{D,PA}} - \underline{\C}_{\text{PA}}.
\end{align}
In Fig.~\ref{fig:MCgapPA_N}, we show how the average gap per complex dimension given by $g_{\text{PA}}$ is always smaller than that deriving from $g_{\text{D,PA}}$. Finally, in Fig.~\ref{fig:PA_SPvsDytso} we show the capacity bounds for a random channel realization, given $\N=10$. As for the TA constraint, it can be seen how the bound $\overline{\C}_{\text{PA}}$ improves significantly upon the upper bounds $\overline{\C}_{\text{D,1}}$ and $\overline{\C}_{\text{D,2}}$ of~\cite{Dytso}.

\section{Conclusion}
\label{S:conclusion}

We derived an upper bound on the channel capacity of multiple-input multiple-output (MIMO) systems affected by fading and subject to peak amplitude constraints at the transmitter. We also introduced two variants of the proposed upper bound. One is used to always ensure a feasible evaluation of the bound and the other improves the performance at low signal-to-noise ratio (SNR). Moreover, we specialized the upper bounds for two particular constraints induced by practical transmitter configurations. The first configuration considers a transmitter employing a single power amplifier and determines a constraint on the norm of the input vector. The other configuration utilizes multiple amplifiers, one per transmitting antenna and it induces a constraint on the peak amplitude of each entry of the input vector. We proved that, for both configurations, the average capacity gap between the upper bounds and the best available lower bound, tends to vanish at high SNR. Moreover, we showed that the capacity gap is limited virtually at any SNR level and for any of the considered MIMO dimensions. We also showed that the presented bounds represent a substantial improvement compared to the previously available upper bounds. Furthermore, we proved that the presented upper bounds are asymptotically tight at high SNR, not only for the considered constraints, but also for any peak amplitude convex constraint, for any channel matrix realization, and any dimension of the MIMO system.

\bibliographystyle{IEEEtran}
\bibliography{bibliofile}

\end{document}